\documentclass{article}
\usepackage{amssymb}

\input{tcilatex}
\begin{document}

\begin{center}
{\LARGE The universal quantum driving force to speed up a quantum
computation --- The unitary quantum dynamics\bigskip }

{\large Xijia Miao\footnote{%
Email: miaoxijia@yahoo.com}}

{\large Somerville, Massachusetts}

{\large Date: February, 2011\bigskip }

{\Large Abstract}
\end{center}

On the basis of the polynomial-time unstructured quantum search process it
is shown that the unitary quantum dynamics in quantum mechanics is the
universal quantum driving force to speed up a quantum computation. This
assertion supports strongly in theory that the unitary quantum dynamics is
the fundamental and universal principle in nature. On the other hand, the
symmetric structure of Hilbert space of a composite quantum system is the
quantum-computing resource that is not owned by classical computation. A new
quantum-computing speedup theory is set up on the basis of the unitary
quantum dynamics in a universal quantum computational model. Both the
unitary quantum dynamics and the symmetric structure and property of the
Hilbert space are mainly responsible for an exponential quantum-computing
speedup for a general efficient quantum algorithm. The inherent importance
for the unitary quantum dynamics to speed up a quantum computation lies in
the unique ability of the unitary quantum dynamics to build the effective
interaction between the symmetric structure of the Hilbert space (or the
fundamental quantum-mechanical principles) and the mathematical symmetric
structure (or the mathematical logic principles) of a problem to be solved
on the quantum system. This unique ability is not owned by the reversible
classical mechanics and the reversible equilibrium-state thermodynamics. It
may result in an essential difference of computational power between quantum
computation and classical computation by combining the symmetric structure
and property of the Hilbert space. In theory and experiment this effective
interaction may be built with the help of the unitary manipulation on the
mathematical-logic functional operations of the problem. The
quantum-computing speedup theory also provides reasonable mechanisms for
exponential quantum-computing speedup for the existing efficient quantum
algorithms that are constructed in the frame of the quantum parallel
principle. These existing quantum algorithms including the
hidden-subgroup-problem quantum algorithms and conventional quantum search
algorithms have the common character that the symmetric structure and
property of the Hilbert space does not have any effective effect on these
quantum algorithms. This could be the main reason why these quantum
algorithms including the efficient ones are quite special and considered to
be semiclassical.\newline
\newline
\newline
{\Large 1. Introduction}

Quantum computation is based on both quantum mechanical principles and
mathematical logic principles. It employs quantum mechanical principles to
study computational science. Its theoretical basis is quantum physics.
Therefore, quantum computation is essentially different from classical
computation based on classical physics. At present its influence may reach
far beyond the quantum computational science. It has already a great
influence on the physical science. It could revolutionize the future
computational science and technology. It will have an extensive influence on
many other science disciplines in future. Quantum computation is mainly
referred to mathematical-sense universal quantum computation. It also may
contain quantum simulation which is usually purely quantum mechanical.
Historically the emergence of quantum computation is related to classical
computation and especially reversible classical computation. This may be
seen in the historical evolution from reversible classical computation to
universal quantum computation. A classical computation obeys the classical
physical laws. It also is required to obey the mathematical-logic
principles. These are the theoretical basis for all digital computers today.
It is well known that a classical computation is irreversible in
mathematical logic, but time evolution process of the classical physical
system to execute a computation is also irreversible. Due to both the
irreversibilities an irreversible classical computation can generate a large
amount of heat in a high-speed computer today [1]. This heating is a severe
limitation for a classical computer to operate in a very high speed. In
order to resolve the heating problem the reversible classical computational
models had been proposed in the middle of the last century [2]. The basic
idea for the reversible computational models is that every computational
step is made mathematical-logically reversible. That the mathematical-logic
operations are made reversible is not the final scheme to resolve thoroughly
the heating problem for a high-speed classical computer, because the time
evolution process of a classical physical system to execute a computation
still could be irreversible. For example, the SWAP operation is reversible
in mathematical logic, but it can be realized directly on an irreversible
classical computer. One might employ an ideal reversible process of
macroscopic thermodynamics [2a] or classical mechanics [2c] to realize a
reversible computation, so that both the mathematical-logic reversibility
and the physical-process reversibility could be achieved in computation at
the same time. But such an ideal physical process is generally hard to
realize in practice. On the other hand, it is well known that according to
quantum mechanics a microscopic physical system obeys naturally the unitary
quantum dynamics, i.e., the Schr\"{o}dinger equation. When a computation is
carried out in a microscopic quantum physical system, the computational
process obeys naturally the unitary quantum dynamics and hence it is
naturally reversible. Thus, the heating problem could be resolved thoroughly
by employing the unitary operators of a microscopic physical system to
realize the reversible mathematical-logic operations in a computational
process. Both the quantum Turing machine model [3] and the universal quantum
Turing machine model [4] use the unitary operators to realize a
computational process. Therefore, there is not any heating problem in both
the computational models.

Now every irreversible mathematical logic operation of a classical
computation is made reversible according to the reversible classical Turing
machine model [2a, 2b]. Then the reversible mathematical logic operation is
replaced with the discrete unitary operation according to the unitary
quantum Turing machine model [3] so that the physical process of the
computation is made unitary, while at the same time keeping every
mathematical logic operation reversible. This shows that both the reversible
classical computational models [2] and the quantum Turing machine models [3]
together may ensure that a computational process obeys the discrete unitary
quantum dynamics (in physical process) and is also compatible with the
mathematical logic principles in a reversible or unitary form (in
computation). However, both the reversible classical computational models
[2] and the quantum Turing machine models [3] are not thought of as
universal quantum computational models.

Since the early 1980s the main research on the quantum computational models
and especially the universal quantum computational models which include the
universal quantum Turing machine [4] and the universal quantum circuit model
[5] has focused on investigation of the computational performance and
especially the computational power. Though the heating problem could be
resolved in the reversible classical computational models [2] and the
quantum Turing machine models [3], it could also be resolved by using other
schemes. For example, by improving cooling techniques or lowering the energy
consuming in computation the heating problem could be resolved as well.
Therefore, the computational performance becomes a key problem to a
universal quantum computer. Whether or not someday a quantum computer can
replace a classical computer is greatly dependent upon its computational
performance. It has been believed extensively [4] that every finitely
realizable physical system can be simulated efficiently by a universal
quantum Turing machine (the quantum Church-Turing thesis). Both a universal
quantum Turing machine model and a universal quantum circuit model are
equivalent to each other in quantum computation [6]. These mean that in
order to investigate the computational performance of a quantum computer one
needs only to investigate the computational performance of a universal
quantum computational model [4, 5]. It also has been believed extensively
[8] that it is essentially difficult to simulate efficiently a quantum
physical system on a classical computer. This implies that a quantum
computational model which is based on the quantum mechanical principles
could not be simulated efficiently by a classical computational model. Both
the reversible computational models [2] and the quantum Turing machine
models [3] can be simulated efficiently by a classical Turing machine.
Therefore, while the reversible computational models are clearly classical,
the quantum Turing machine models still could be classical computationally
in the sense that they can be efficiently simulated by a classical Turing
machine, and they are not a universal quantum computational model. A
universal quantum computational model [4, 5, 6, 7] is based on quantum
mechanics. According to quantum mechanics [9] a closed quantum system is
described completely by the quantum states of the Hilbert state space of the
quantum system and its time evolution process obeys the unitary quantum
dynamics. Because a universal quantum computational model is based on
quantum mechanics, the time evolution process of the quantum system to
execute a quantum computation obeys the unitary quantum dynamics, while the
quantum system itself is described completely by its quantum states.
Therefore, a universal quantum computational model still obeys the unitary
quantum dynamics and may make use of a variety of fundamental properties of
quantum states of the quantum system in a computation. These fundamental
quantum-state properties include the quantum superposition principle, the
quantum coherence interference effects, and also include the properties of
the special quantum states such as the quantum entanglement states, the
multiple-quantum coherence mixed states, and so on. It seems that there is
not any difference between a universal quantum computational model and
quantum mechanics. In fact, there must be a constraint on a universal
quantum computational model so that it can be thought of as a computational
model instead of quantum mechanics. This constraint is that a universal
quantum computational model must be compatible with mathematical-logic
principles, as emphasized explicitly in Refs. [21a, 29]. This leads to that
the interaction becomes inevitable between the mathematical logic principles
that a mathematical or computational problem obeys and the fundamental
quantum-mechanical principles that a universal quantum computer obeys.
Mathematical-logic principles that a problem obeys are naturally carried out
by the fundamental mathematical-logic operational rules in a computation [2,
3, 4, 5, 6, 7, 31]. They neither belong to the classical physics nor the
quantum physics, that is, they are independent of both the classical and
quantum physics. Therefore, a universal quantum computational model must
obey the three fundamental principles (or properties): $(i)$ the physical
process of a quantum computation obeys the unitary quantum dynamics, $(ii)$
a quantum computation is compatible with the mathematical logic principles
in a reversible or unitary form, meaning that the quantum computation obeys
the mathematical logic principles in a reversible or unitary form, $(iii)$
the quantum system to execute a quantum computation is described completely
by its quantum states and the quantum computation allows to make use of a
variety of fundamental quantum-state properties of the quantum system. Some
of these fundamental principles (or properties), that is, the unitary
quantum dynamics, the mathematical-logic principles, and the fundamental
quantum-state properties are believed to be mainly responsible for the
computational power of a universal quantum computer that can outperform
essentially a classical computer. Since the mathematical-logic principles
(including the mathematical symmetric structures, etc., of course) are
independent of a detailed physical computational model, they are considered
not to have an essential effect on the computational power of a quantum
computer. A similar viewpoint [4, 5] that pure mathematics does not affect
essentially computational performance of a computer also was pointed out in
the past. Therefore, only the fundamental principles (or properties) in
quantum mechanics may have a possibility to make an essential influence on
the computational power of a universal quantum computer and especially to be
responsible for an exponential quantum-computing speedup. According to
quantum mechanics there may be two essentially different quantum-computing
speedup theories. One is based on the fundamental quantum-state properties.
This is the earliest quantum-computing speedup theory ----- the quantum
parallel principle [4] which is based on the quantum-state superposition
principle in quantum mechanics. Another is based on the unitary quantum
dynamics [21, 32]. Though mathematical-logic principles are not essentially
responsible for a quantum-computing speedup, they play a key role in judging
which one of the two quantum-computing speedup theories to be responsible
essentially for a quantum-computing speedup. The unitary quantum dynamics
has been overlooked in studying mechanisms of quantum-computing speedup in
the past decades. Its inherent importance to speed up a quantum computation
was not revealed until these extraordinarily fast quantum algorithms [21,
32] are discovered by the present author in the early 2000s. It is this
inherent importance (See the comment [33]) that ultimately leads the present
author to put the unitary quantum dynamics at the center of the universal
quantum computational models and suggest it as the fundamental and universal
quantum-mechanical principle in nature.

In the past decades the computational performance of a quantum computer has
been extensively investigated. It was shown [10, 3] that the reversible
classical computational models and the quantum Turing machine models can be
as powerful as the irreversible classical computational model. Feynman [8]
had paid attention to the known fact that simulating a quantum system is
essentially difficult on a classical computer, and more importantly he
suggested that this difficulty could be avoided with the help of the
principles of quantum mechanics. Feynman$^{\prime }$s idea implies that a
quantum computer might be more powerful than a classical computer. An
important and essential advance came soon after the universal quantum Turing
machine model [4] was set up. Deutsch proposed for the first time a quantum
algorithm according to the quantum parallel principle [4] and then Deutsch
and Jozsa [11] generalized the quantum algorithm to show that a quantum
computer indeed can solve a problem in polynomial time, while a
deterministic classical computer can not solve the same problem in
polynomial time. This result also was further confirmed and developed [12].
This is the first remarkable example that the quantum parallel principle
based on the superposition principle of quantum mechanics plays an important
role for a quantum algorithm outperforming essentially its classical
counterpart. Soon after the remarkable work [11] several important quantum
algorithms [7, 13, 14] were discovered also based on the quantum parallel
principle. They further show that a quantum computer even can outperform
essentially a probabilistic classical computer. Since then, the quantum
parallel principle is considered as a fundamental rule to guide the design
of a quantum algorithm. Many quantum algorithms [11, 12, 13, 14, 15, 16]
have been discovered that can outperform essentially their classical
counterparts under the guidance of the quantum parallel principle in the
past two decades. Among these quantum algorithms the polynomial-time quantum
algorithms for the prime factorization and discrete logarithms [14] are one
of the important applications of the quantum parallel principle. These
quantum algorithms show that the quantum parallel principle indeed makes a
great contribution to development of the quantum computational science in
the past two decades. However, at present it is believed that the quantum
parallel principle is not enough powerful to solve many important problems
such as the graph isomorphism problem and most non-Abelian hidden subgroup
problems [16]. It is believed extensively that these non-Abelian hidden
subgroup problems are not harder than the NP-complete problems [17] in
computational complexity.

Since the oracle-based or black-box-based quantum algorithms [11, 12, 7, 13]
were first formulated, a large number of oracle-based quantum algorithms
have been discovered to solve the specific problems on a quantum computer.
An unstructured search problem could be the most important one of these
problems. An unstructured quantum search algorithm has an extensive
application in practice. In particular, it may be used to solve an
NP-complete problem. It is well known that most important problems in
computational science are either polynomial-time or NP-complete. The
oracle-based quantum search algorithms have been investigated extensively in
the past decades. The conventional (unstructured) quantum search algorithm
(i.e., the Grover$^{\prime }s$ quantum search algorithm [23]) is the first
oracle-based quantum search algorithm with a square speedup. It is based on
the amplitude-amplification mechanism [23, 24]. Another important quantum
search algorithm is the adiabatic quantum search algorithm [25]. It is based
on the\ continuous-time adiabatic state-transfer mechanism in a
space-independent quantum system [34]. Though these quantum search
algorithms are discovered also under the guidance of the quantum parallel
principle, their computational power is far more weak than an exponential
speedup in worst case. Actually, it has been shown that the conventional
quantum search algorithm [23, 24]\ is square speedup and moreover this
square speedup is optimal [26], and the adiabatic quantum search algorithm
is also square speedup in worst case [27]. More generally, it has been shown
that any oracle-based quantum algorithm to compute a total Boolean function
can achieve only a polynomial speedup over its classical counterpart [28].
Therefore, all these oracle-based quantum search algorithms are not enough
powerful to solve efficiently the NP-complete problems.

The quantum simulation was initiated by the Feynman$^{\prime }$s work [8].
When the mathematical-logic principles are not considered explicitly, the
universal quantum circuit model [5] is reduced to the quantum simulator. A
usual quantum simulation usually does not consider explicitly the
mathematical-logic principles. Thus, it is usually thought of as a purely
quantum-mechanical process instead of a mathematical-sense quantum
computation. This results in that a number of important problems in quantum
computation can not be owned by a conventional quantum simulation. Many
fundamental and essential problems of quantum computation can not be solved
by studying alone a conventional quantum simulation. Based on the Feynman$%
^{\prime }$s ideas several important works [18] on the quantum simulation
have been developed in detail. These works show that many quantum systems
may be simulated efficiently on a quantum simulator, while they could not be
simulated efficiently on a classical computer or it has not yet been found
that there are efficient algorithms to simulate these quantum systems on a
classical computer. It has been shown [18d] that the quantum parallel
principle still plays an important role for the quantum simulation
outperforming the classical simulation. The Trotter-Suzuki formalism [19,
20] could provide a unified frame to describe both a conventional quantum
simulation and a classical simulation. It has been shown that it is hard to
solve an unstructured search problem by the quantum simulation [55]. Though
the special unitary quantum simulation may be used to solve an unstructured
search problem in a quantum spin system (or ensemble) [21], it has not yet
been proven that it is both efficient and scale. It also has been shown [22]
that most unitary transformations can not be simulated efficiently on a
quantum computer. Hence the quantum simulation could not be enough powerful
to solve efficiently an unstructured search problem.

It has been believed extensively that an NP-complete problem can not be
solved efficiently on a classical computer [17]. It has turned out that a
quantum computer is more powerful than a classical computer. Then one of the
most important problems in the quantum computational science is to answer
the question: whether or not a universal quantum computer is capable of
solving an NP-complete problem in polynomial time. As described in the
previous paragraphs, a number of efficient quantum algorithms have been
discovered in the past two decades and most of them are discovered in the
frame of the quantum parallel principle. A detailed review for a variety of
quantum algorithms including those mentioned above, the topological quantum
algorithms, the random-walk-based quantum algorithms, etc., may be seen in
recent review papers [15d, 15e]. However, no one of these quantum algorithms
can show that a universal quantum computer is able to solve an NP-complete
problem in polynomial time.

Indeed, in the past decades the quantum parallel principle achieves a great
success in guiding construction of the efficient quantum algorithms, but
after examining these existing quantum algorithms, as described above, one
may find that only some special quantum algorithms can achieve an
exponential quantum-computing speedup, which include mainly the Abelian
hidden-subgroup-problem (HSP) quantum algorithms [11 -- 15] and some
non-Abelian HSP quantum algorithms [16, 15d, 15e]. This shows that the
quantum parallel principle is greatly limited. A question therefore arises
whether or not the quantum parallel principle is really responsible for an
essential quantum-computing speedup in quantum computation. This question
will be answered in the paper. This paper is mainly devoted to studying
mechanisms of quantum-computing speedup and especially exponential
quantum-computing speedup in quantum computation. It investigates how the
fundamental quantum-mechanical principles and properties affect
computational power of a universal quantum computer. The research in the
paper is based on the present author$^{\prime }$s works in the past decade
and particularly the polynomial-time unstructured quantum search processes
[56].\newline
\newline
{\Large 2. The Symmetric structure and property of direct-product Hilbert
space and the polynomial-time unstructured quantum search processes}

A quantum algorithm may be constructed generally in the frame of a universal
quantum computational model, but it may be realized in various quantum
systems. It is also possible that a quantum system could be better to
realize a quantum computation than other quantum systems. It seems that
difference in computational power for these quantum systems could not be
essential according to the quantum Church-Turing thesis [4]. However, a
universal quantum computational model [4, 5, 6, 7] generally does not
consider explicitly whether or not the symmetric structure and property of
the Hilbert space of a quantum system to execute a quantum computation may
affect essentially quantum computational performance. Here the Hilbert space
may be referred to the Hilbert state space or its corresponding operator
space of a quantum system (the operator space also is often called the
Liouville operator space in a quantum ensemble). Obviously, a universal
quantum computational model allows ones to make use of the symmetric
structure and property of the Hilbert space to improve the quantum
computational performance [21, 29]. For example, the symmetrical structure
and property of the multiple-quantum operator algebra space [35] has been
used to simplify the unitary quantum simulations in quantum spin systems or
ensembles (See also the present author$^{\prime }$s early works [36, 37]).
The symmetric structure of the Hilbert space of a quantum system is another
fundamental attribute of quantum mechanics which is different from the
unitary quantum dynamics and the fundamental quantum-state properties. It is
the quantum-computing resource that is not owned by classical computation.
It could affect essentially quantum computational performance. In quantum
mechanics [9] this fundamental attribute is closely related to the basic
postulate that the Hilbert space of a composite quantum system is a direct
product of the Hilbert spaces of the component systems of the composite
system. Here it must be emphasized that symmetric structure of the Hilbert
space of a quantum system is different in concept from pure mathematical
symmetric structure of a problem to be solved. The former belongs to quantum
physics, while the latter is of pure mathematics and does not have an
essential effect on quantum computational performance. Then in quantum
computation there are the two essentially different symmetric structures.
One of which is of quantum system (or quantum computer) and another of a
problem to be solved [21a]. In quantum computation one must explicitly
distinguish the two essentially different symmetric structures from each
other. As a typical instance, in quantum computation a hidden subgroup (HS)
problem [13, 14, 15, 16] may be defined according to the symmetric structure
of a specific group, e.g., an Abelian group or a non-abelian group, but it
is still a pure mathematical problem. Its group symmetric structure is
independent of any detailed physical computational model. Hence it is
different from any symmetric structure of the Hilbert space of the quantum
system that is used as a quantum computer to solve the HS problem. It does
not affect essentially quantum computational performance of the quantum
computer. Another typical example is a structured search problem in quantum
computation. The mathematical symmetric structure of a structured search
problem is different from any symmetric structure of the Hilbert space of
the quantum system used to solve the search problem. It does not affect
essentially the quantum-searching speedup. Thus, just like an unstructured
quantum search algorithm, a structured quantum search algorithm only can
achieve a square speedup with respect to its classical counterpart [38].
Researchers could confuse the mathematical symmetric structure of a
mathematical or computational problem with that one of the Hilbert space of
the quantum system used to solve the problem. Sometimes this is because the
two symmetric structures are in the same Hilbert space in quantum
computation. A typical example may be seen in the conventional unstructured
quantum search algorithm [23, 24], where the search space of the
unstructured search problem is directly taken as the Hilbert space of the $%
n- $qubit quantum system used to solve the search problem. It has been
extensively investigated [13, 14, 15, 16] how a quantum computer could solve
efficiently the HS problems that have the specific group symmetric
structures. However, here it is investigated how the symmetric structure and
property of the Hilbert space of a quantum system may affect or even improve
essentially computational performance of any quantum computational process
running in the quantum system to solve a problem that could have some purely
mathematical symmetric structure. It is clear that the two investigations
are two completely different things.

A conventional quantum search algorithm usually works in an $n-$qubit spin
system. Such a spin system has a typical tensor product (or direct product)
symmetric structure in its Hilbert space. It has been shown that such a
tensor-product symmetric structure is not necessary to achieve a square
speedup for a conventional quantum search algorithm [39]. That is, the
square speedup has nothing to do with the symmetric structure of the
tensor-product Hilbert space of the $n-$qubit spin system. However, it is
shown below that the symmetric structure of the tensor-product Hilbert space
is necessary for a polynomial-time unstructured quantum search process. In a
usual quantum search algorithm the search space for an unstructured search
problem is usually taken as the Hilbert space of a quantum system. From the
point of view of pure mathematics the search space is unstructured, while
from the point of view of quantum mechanics the Hilbert space that serves as
the search space may be structured or unstructured, which is dependent on
the quantum system. Here researchers should pay attention to the fact that
the symmetric structure of the search space may be different from that one
of the Hilbert space. The former originates from the search problem or the
search algorithm, while the latter is of the quantum system (or the quantum
computer). However, if one ignores any possible symmetric structure of the
Hilbert space, then the Hilbert space may be treated as an unstructured
space even if it is a tensor-product Hilbert space of an $n-$qubit spin
system. This means that in quantum computation there is no way to reduce the
exponentially large search space [54] to a polynomially small search space
[54] for an unstructured search problem and there is no way to reduce the
unstructured quantum search process to a structured one unless the symmetric
structure and property of the Hilbert space is considered explicitly. The
conventional quantum search algorithms [23, 24] and the adiabatic quantum
search algorithms [25] do not consider explicitly the symmetric structure
and property of the Hilbert space. They have an exponentially large search
space, resulting in that they cannot achieve an exponential
quantum-searching speedup. In fact, a conventional quantum search algorithm
[23, 24] can achieve only a square speedup\ [26] and the adiabatic quantum
search algorithm [25] also achieves only a square speedup in the worst case
[27]. Therefore, there is no way to improve essentially these quantum search
algorithms (See, Refs. [26, 27]) unless the symmetric structure and property
of the Hilbert space is considered explicitly. Furthermore, it will be seen
later that even if the symmetric structure and property of the Hilbert space
of an $n-$qubit spin system is considered explicitly, these usual quantum
search algorithms [23, 24, 25] that work in the spin system could not be
improved essentially.

If there exists the proper symmetric structure and property in the Hilbert
space, then there exists a possibility that the symmetric structure and
property of the Hilbert space could affect essentially an unstructured
quantum search process and could be exploited to speed up the quantum search
process [21, 29]. One might image intuitively that the unstructured search
process could be reduced to some structured quantum search process due to
the effect of the symmetric structure and property of the Hilbert space on
the unstructured search process, resulting in that the unstructured search
process is sped up. The problem is that the symmetric structure and property
of the Hilbert space is independent of any mathematical logic principles
including the mathematical symmetric structure of the unstructured search
problem. How can the symmetric structure and property of the Hilbert space
affect the unstructured search process? There must exist the
quantum-mechanical principle to bring together both the symmetric structure
and property of the Hilbert space and the mathematical logic principles of
the search problem. Otherwise the symmetric structure and property of the
Hilbert space could not affect essentially the quantum search process. This
fundamental quantum-mechanical principle is the unitary quantum dynamics.
The importance inherent in the quantum-computing speedup for the unitary
quantum dynamics is that the unitary quantum dynamics is able to build the
effective interaction between the symmetric structure and property of the
Hilbert space of a quantum system and the mathematical logic principles of a
problem to be solved on the quantum system. Whether or not this possibility
that the symmetric structure and property of the Hilbert space could affect
essentially an unstructured search process can become real is completely
dependent on the unitary quantum dynamics. Therefore, there are the two
fundamental quantum-mechanical principles to help an unstructured quantum
search process to bypass the square speedup limitation. One of which is the
unitary quantum dynamics in time and space. Another is the symmetric
structure and property of the Hilbert space of a quantum system. In the past
decade around the present author has investigated extensively how the two
fundamental quantum-mechanical principles can improve essentially the
computational performance of a quantum search process. These two fundamental
quantum-mechanical principles lead to that there are polynomial-time
unstructured quantum search processes in quantum computation [56]. These
quantum search processes may be used to solve the $NP-$complete problems in
polynomial time, indicating that in computational complexity there is the
relation $\mathbf{NP=P}$ on a universal quantum computer. A polynomial-time
quantum search process [56] consists of the two different parts in
structure. The first part [21, 29, 37, 35] is to realize dynamically an
efficient reduction from the exponentially large Hilbert space that serves
as the unstructured search space to the polynomially small state subspace
(or state subset) of the Hilbert space in which the quantum states carry the
information of the component states of the marked state of the unstructured
search problem. This part is realized on the basis of both the unitary
quantum dynamics and the symmetric structure and property of the
tensor-product Hilbert space. The second part is to realize an exponential
quantum-state-difference amplification. It is really an inverse process of
the unitary dynamical state-locking process [30]. An exponential
quantum-state-difference amplification usually could be realized in the
time- and space-dependent quantum system of a single atom motioning in time
and space. Therefore, the second part is realized on the basis of the
unitary quantum dynamics in time and space.

In a polynomial-time quantum search process the symmetric structure and
property of the tensor-product Hilbert space of an $n-$qubit (or more
generally $n-$partite) composite quantum system has to be considered
explicitly, while this consideration is not necessary in a conventional
quantum search algorithm. This leads to that there is an essential
difference in theory between a polynomial-time quantum search process and
both the conventional quantum search algorithms [23, 24] and the adiabatic
quantum search algorithm [25]. A usual quantum search algorithm [23, 24] may
be generally written as%
\begin{equation}
|\Psi _{f}^{S}\rangle =U_{k}C_{S}(\theta _{k})U_{k-1}C_{S}(\theta
_{k-1})...U_{1}C_{S}(\theta _{1})|\Psi _{0}\rangle \rightarrow |S\rangle , 
\tag{1}
\end{equation}%
where $C_{S}(\theta _{k})$ is a general reversible oracle operation
selectively applying to the solution state $|S\rangle $ to the unstructured
search problem. The solution to the search problem can be obtained directly
from the solution state $|S\rangle $. The solution state also is called the
marked state of the search problem. An adiabatic quantum search algorithm
also may be efficiently reduced to the unitary quantum circuit (1) [27a].
This quantum search algorithm (1) consists of the two types of unitary
operations. One of which is the oracle operations $\{C_{S}(\theta _{k})\},$
here each oracle operation may be implemented by the reversible operation
sequence $C_{S}(\theta _{k})=U_{f}V(\theta _{k})U_{f}$ consisting of the two
oracle functional operations $U_{f}$ and a conditional phase shift operation 
$V(\theta _{k})$ [21a]. Another consists of the known unitary operations $%
\{U_{k}\}$. The reversible oracle operations $\{C_{S}(\theta _{k})\}$ in (1)
are independent on any symmetric structure and property of the Hilbert space
that serves as the search space. The known unitary operations $\{U_{k}\}$
could be constructed according to the symmetric structure of the Hilbert
space, but they also may be built up without considering the symmetric
structure of the Hilbert space [39, 24]. Therefore, it may be thought that
the quantum search algorithm is independent on any symmetric structure and
property of the Hilbert space. Then this means that the search space of the
quantum search algorithm is not essentially different from the unstructured
search space of the search problem, even though it is just taken as the
tensor-product Hilbert space of the $n-$qubit spin system. The final state $%
|\Psi _{f}^{S}\rangle $ in (1) is required to be close to the solution state 
$|S\rangle $ or to contain the solution state in a high probability close to
100\%, so that the solution state can be found by quantum measuring the
final state $|\Psi _{f}^{S}\rangle $. This directly leads to that number of
the oracle operations in (1) (or equivalently the oracle functional
operations) needs to take $O(\sqrt{N})$ [23, 24]. On the other hand, a
classical search algorithm needs to evaluate the oracle function $O(N)$
times to determine the solution $S$ that corresponds to the solution state $%
|S\rangle $ in a quantum search algorithm. Therefore, in comparison with the
classical search algorithm the quantum search algorithm (1) achieves only a
square speedup. This is the square speedup limitation on a conventional
quantum search algorithm. It has been shown that the quantum search
algorithm (1) obeys generally the square speedup limitation [26, 27], here
the oracle operation in (1) may be taken as a general one $C_{S}(\theta
_{k}) $ with variable angle $\theta _{k}$ or the special one $C_{S}(\pi )$
with the fixed angle $\pi .$

In a polynomial-time quantum search process the efficient reduction from the
exponentially large search space to some polynomially small subspace of the
Hilbert space is carried out by the unitary quantum circuit: 
\begin{equation}
|\Psi _{f}\rangle =U_{k}C_{S}(\theta _{k})U_{k-1}C_{S}(\theta
_{k-1})...U_{1}C_{S}(\theta _{1})|\Psi _{0}\rangle \rightarrow \sum_{L\neq
S}B_{L}^{S}|L\rangle .  \tag{2}
\end{equation}%
This quantum circuit is completely the same as that one of (1) in form.
However, in the final superposition state $|\Psi _{f}\rangle $ of the
quantum circuit (2) there is not the solution state $|S\rangle $ or the
solution state $|S\rangle $ can be neglected with respect to the other
states. This is essentially different from the usual quantum search
algorithm (1). This is also a necessary condition to realize a reduction
from the exponentially large search space to a polynomially small one. Only
when this necessary condition is met, can the exponentially large search
space be possibly reduced to a polynomially small subspace. In fact, if
there is the solution state $|S\rangle $ in the final state $|\Psi
_{f}\rangle ,$ then the search space is always exponentially large, because
the solution state $|S\rangle $ can be any quantum state of the
exponentially large search space. Because the amplitude of the solution
state $|S\rangle $ is zero or negligible in the final state $|\Psi
_{f}\rangle $, there is not the square speedup limitation on the quantum
circuit (2)\ that is used to realize the search-space reduction. Therefore,
the quantum circuit (2) could be realized efficiently. Because there is not
the solution state $|S\rangle $ in the final state, the original search
space of the unstructured search problem and its symmetric structure
disappear. This is a necessary condition for a quantum search process to be
able to solve an unstructured search problem in polynomial time. The usual
quantum search algorithm (1) still works in the original unstructured search
space. It is not able to solve an unstructured search problem in polynomial
time. Now one may ask whether or not the information of the solution state $%
|S\rangle $ also disappears. Actually, the information of the solution state 
$|S\rangle $ is transferred to those quantum states $\{|L\rangle \}$ with $%
|L\rangle \neq |S\rangle $ (or their amplitudes $\{B_{L}^{S}\})$ of the
Hilbert space of the quantum system that serves as the search space. It is
clear that the final state $|\Psi _{f}\rangle $ does not carry the
information of the solution state as a whole but the information of the
component states of the solution state. By comparing the two quantum
circuits (1) and (2) with each other one can see that there exist the two
extreme cases. One extreme case is that the final state $|\Psi
_{f}^{S}\rangle $ in (1) is completely the solution state $|S\rangle $ for a
usual quantum search algorithm. Another is that the final state $|\Psi
_{f}\rangle $ in (2) does not contain the solution state $|S\rangle $ at all
for the search-space reduction in a polynomial-time quantum search process.
This result is very surprising. For a long time researchers have made
continuously a great effort to amplify maximally the amplitude of the
solution state $|S\rangle $ with the minimum number of the oracle operations
in a quantum search algorithm. However, the successful direction to solve
efficiently the unstructured search problem may be the opposite direction to
the amplitude amplification of the solution state.

In a polynomial-time quantum search process the original unstructured search
space and its symmetric structure are first eliminated so that they cannot
affect effectively the quantum search process. At the same time the
information of the component states of the solution state is transferred to
the tensor-product Hilbert space of the quantum system. Here the
tensor-product symmetric structure of the Hilbert space is of crucial
importance, because without the tensor-product symmetric structure the
information of the component states will not be smoothly transferred into
the Hilbert space. Because the information treatment for the component
states is the main task in a polynomial-time quantum search process, the
unitary operations, excitations, and processes in the component systems of
the quantum system are more important in the quantum search process. This
could be different from a usual quantum search algorithm, where the
information treatment for a quantum state as a whole is the main task.
Unitarily manipulating the oracle operations (or the oracle functional
operations) [21, 29] is necessary to realize the transfer of information of
the component states of the solution state from the original search space
into the Hilbert space of the quantum system. This is essentially different
from a usual quantum search algorithm [23, 24, 25]. Unitarily manipulating a
functional operation which is not an oracle operation is also very important
in quantum computing [40]. It is generally required in quantum computation
that unitary manipulation and control obey the mathematical logic principles
of a problem such as the unstructured search problem. This is different from
conventional unitary manipulation and control in quantum mechanics. The
symmetric structure and property of the Hilbert space must be considered
explicitly in the unitary manipulation in a polynomial-time quantum search
process, resulting in that the unitary manipulation is performed mainly in
the component systems of the quantum system. Therefore, although both the
quantum circuits (1) and (2) are the same in form, they are essentially
different from each other. On the one hand, because the mathematical logic
principles have to be obeyed, both the quantum circuits (1) and (2) must
contain the same reversible oracle functional operations. On the other hand,
the known unitary operations $\{U_{k}\}$ in (2) which are purely
quantum-mechanical can not be taken as arbitrary unitary operators, because
they are used for the purposeful unitary manipulation on the oracle
operations, and thus, they are unlike those known unitary operations in (1)
which are also purely quantum-mechanical. It is this purposeful unitary
manipulation that is used to suppress or even cancel the effect of the
original search space and its symmetric structure on the quantum search
process and at the same time realize the transfer of information of the
component states. \textit{Here the mutual cooperation between the reversible
oracle operations and the known unitary operations in the frame of the
unitary quantum dynamics reflects the importance of cooperation between the
mathematical logic principles of a problem to be solved and the
quantum-mechanical principles in quantum computation}.

Though the original search space and its symmetric structure disappear and
there does not exist the square speedup limitation on the quantum circuit
(2) of the search-space reduction, these could not guarantee that the new
search space, i.e., the Hilbert space is not exponentially large for the
unstructured search problem. For example, the solution state could disappear
apparently, and it could be changed merely from one form to another. At
present it is unclear in what conditions the Hilbert space is not
exponentially large for the unstructured search problem. However, it is
essential for the transfer of information of the component states of the
solution state from the original search space into the Hilbert space. This
is because the Hilbert space has a tensor-product symmetric structure. The
symmetric structure and property of the Hilbert space could be exploited to
suppress or even cancel the effect of the symmetric structure of the
original search space on the quantum search process. \textit{More
importantly one may make use of the symmetric structure and property of the
tensor-product Hilbert space to help solving the unstructured search problem}
[21, 37, 29]. Information of all these component states of the solution
state contains the information of the solution state as a whole. If the
information of all these component states is extracted from the Hilbert
space, then the solution state can be determined completely. Therefore,
extracting the information of all these component states is really
equivalent to solving the unstructured search problem. Around ten years ago
[21] the present author attempted to extract directly the information of
these component states from the Hilbert space of an $n-$qubit spin system.
The problem for the scheme is that either the information transfer is
inefficient (the main one) or the information of all these component states
is distributed in the whole Hilbert space which is exponentially large,
resulting in that it is hard to extract the information or it needs to take
an exponential time to extract the information. Thus, this scheme may be
useful only to solve a small-scale search problem [21]. The key strategy to
solve the problem is to make use of the symmetric structure and property of
the Hilbert space to help extraction of the information of these component
states [21, 37, 29]. According to this strategy a crucial method to solve
the problem is with the aid of the symmetric structure and property of the
multiple-quantum operator spaces of a spin system [21a] which also is a
Hilbert operator space. With the help of the symmetric structure and
property the information of the component states of the solution state is
dynamically transferred to a small state subspace (or state subset) or even
a polynomially small state subspace (or state subset) of the Hilbert space,
and at the same time the solution state disappears. Here the polynomially
small subspace stores only the information of the component states of the
solution state and it does not contain the information of the solution state
as a whole. After this information transfer it may be thought that the
original unstructured search space is reduced to the polynomially small
subspace of the Hilbert space. Dynamically this reduction may be efficiently
realized by unitarily manipulating the oracle operations in the component
systems of the quantum system [21, 29]. Here emphasize that dynamically the
unitary manipulation must be explicitly applied to the component states of
the solution state, so that the information transfer can be realized at the
same time. This efficient reduction is one of the two key steps to solve the
unstructured search problem in polynomial time. It simplifies greatly the
extraction of the information of the component states. However, it is still
hard to solve the unstructured search problem. This is because amplitude of
the quantum state carrying the information of the component states is
exponentially small in the polynomially small subspace, resulting in that
extraction of the information is still hard. The amplitude is even
exponentially smaller than the counterpart in a usual quantum search
algorithm. This could be the price for the efficient reduction from the
original unstructured search space to the polynomially small subspace of the
Hilbert space. In order to extract efficiently the information of the
component states in the polynomially small subspace one needs to use the
exponential quantum-state-difference amplification [30, 56], which the
present author spends the last five years to set up. This exponential
quantum-state-difference amplification is extremely powerful. It is another
key step to solve the unstructured search problem in polynomial time.

As stated before, there are two fundamental quantum-mechanical principles to
help an unstructured quantum search process to bypass the square speedup
limitation. One of which is the symmetric structure and property of the
tensor-product Hilbert space of a quantum system. It has been discussed in
detail above. Another is the unitary quantum dynamics in time and space. A
quantum system like a spin system is simpler to realize a quantum
computation. Its time evolution process may be described simply by a
space-independent unitary quantum dynamics. Today most efficient quantum
algorithms are constructed based on the quantum system. From the point of
view of unitary manipulation such a quantum system has only one manipulating
freedom degree of the internal motion of the system. On the other hand, a
quantum system such as a single atom motioning in time and space has not
only the internal (electron or spin) motion but also the center-of-mass
motion in space. Its time evolution process has to be described by a time-
and space-dependent unitary quantum dynamics. Both the polynomial-time
unstructured quantum search processes [56] and the reversible and unitary
halting protocol [30] are realized in a time- and space-dependent quantum
system. Such a quantum system is far more complicated than a spin system,
but it may be much more useful in the unitary manipulation to realize the
unitary dynamical state-locking process [30, 41, 42, 43] and the exponential
quantum-state-difference amplification [30, 56]. From the point of view of
unitary manipulation a time- and space-dependent quantum system may have two
or more independent manipulating freedom degrees. In particular, a single
atom motioning in time and space is one of the simplest time- and
space-dependent quantum systems. It has two independent manipulating freedom
degrees. One of which is the atomic internal motion and another the atomic
center-of-mass motion. A single atom that has both the independent
manipulating freedom degrees may be used to realize the quantum-state-level
mutual cooperation between both the atomic internal and center-of-mass
motions. Here realization for the quantum-state-level mutual cooperation
needs to manipulate unitarily the discrete atomic internal motion, the
atomic center-of-mass motion, and the coupling between the atomic internal
and center-of-mass motions [30, 41, 42, 43]. \textit{Such
quantum-state-level mutual cooperation plays the key role in realizing the
unitary dynamical state-locking process} [30]. In general, a unitary
dynamical state-locking process may be defined simply and intuitively as a
unitary process that transforms simultaneously two (or more) very
distinguishable quantum states (e.g., a pair of orthogonal states) to two
(or more) indistinguishable quantum states (e.g., a pair of non-orthogonal
states) whose difference may be arbitrarily small. A unitary dynamical
state-locking process may be used to realize the reversible and unitary
halting protocol [30]. The essential difference between a unitary dynamical
state-locking process and a conventional unitary process (or operation) may
be seen intuitively through the following typical instance. Consider that a
single atom in a harmonic potential field is in the product state $%
|s_{k}\rangle |\Psi (x,t_{0})\rangle .$ Here the discrete atomic internal
state $|s_{k}\rangle $ may be either $|0\rangle $ or $|1\rangle $ and the
atomic center-of-mass motional state $|\Psi (x,t_{0})\rangle $ may be the
ground state of a harmonic oscillator, which is a Gaussian wave-packet
state. Now the two orthogonal internal states $|0\rangle $ and $|1\rangle $
may be transformed to the desired internal state $|0\rangle $ simultaneously
in a probability close to unity by the unitary dynamical state-locking
process $U_{DSL}$ in this single-atomic system [30, 56],%
\begin{equation}
U_{DSL}|s_{k}\rangle |\Psi (x,t_{0})\rangle \rightarrow \{\rho
(s_{k})|0\rangle +\exp [i\gamma (s_{k})]\sqrt{1-|\rho (s_{k})|^{2}}|1\rangle
\}|\Psi (x,t_{0})\rangle ,  \tag{3}
\end{equation}%
where $|\rho (s_{k})|^{2}$ is the probability that the internal state $%
|s_{k}\rangle $ with $s_{k}=0$ or $1$ is changed to the desired state $%
|0\rangle $ and $\exp [i\gamma (s_{k})]$ is a phase factor. The unitary
dynamical state-locking process (3) may be realized efficiently. The minimum
one of the two different probability values $\{|\rho (s_{k})|^{2}\}$ in (3)
is defined as $P_{\min }=\min \{|\rho (0)|^{2},|\rho (1)|^{2}\}.$ The
unitary dynamical state-locking process is essentially different from a
usual unitary process (or operation) in that the minimum probability $%
P_{\min }$ $(P_{\min }<1)$ can be made infinitely close to unity without
destroying the unitarity of the process [30, 56]! This extremely important
property leads to that a unitary dynamical state-locking process and its
inverse process together can realize an exponential quantum-state-difference
amplification and hence an efficient quantum search process becomes possible.

A quantum-state-difference amplification is just the inverse process of a
unitary dynamical state-locking process such as (3). An exponential
quantum-state-difference amplification means that the inverse process of a
unitary dynamical state-locking process such as (3) can be realized in
polynomial time even when the difference between both the non-orthogonal
states on the right-hand side of (3) is exponentially small. As a typical
example, on the right-hand side of (3) both the non-orthogonal states $%
|0\rangle $ and $A(|0\rangle +2^{-n}|1\rangle )$ (the normalization constant 
$A\thickapprox 1$ for the qubit number $n>>1$) whose difference is
exponentially small may be transformed simultaneously to the two orthogonal
states $|0\rangle $ and $|1\rangle $ by the exponential
quantum-state-difference amplification of (3) in polynomial time,
respectively. This illustrates that an exponential quantum-state-difference
amplification is extremely powerful to distinguish two non-orthogonal states
unambiguously. Can an exponential quantum-state-difference amplification
help to solve efficiently an unstructured search problem? Consider the two
non-orthogonal states $|\Psi _{k}^{S}\rangle $ and $|\Psi _{k}^{0}\rangle $
[26b, 26c] that are created via the quantum circuit (1) of a usual quantum
search algorithm and its modified version without containing any oracle
operations, respectively,%
\[
|\Psi _{k}^{S}\rangle =U_{k}C_{S}(\theta _{k})U_{k-1}C_{S}(\theta
_{k-1})...U_{1}C_{S}(\theta _{1})|\Psi _{0}\rangle , 
\]%
\[
|\Psi _{k}^{0}\rangle =U_{k}U_{k-1}...U_{1}|\Psi _{0}\rangle , 
\]%
here the oracle number $k=$poly($n$). Clearly only the state $|\Psi
_{k}^{S}\rangle $ contains the information of the solution state $|S\rangle
. $ In most cases both the known state $|\Psi _{k}^{0}\rangle $ and the
unknown state $|\Psi _{k}^{S}\rangle $ have an exponentially small
difference [26]. If both the non-orthogonal states $|\Psi _{k}^{0}\rangle $
and $|\Psi _{k}^{S}\rangle $ could be transformed to their corresponding
orthogonal states by an exponential quantum-state-difference amplification
in polynomial time, then one would be able to distinguish unambiguously the
state $|\Psi _{k}^{S}\rangle $ from the state $|\Psi _{k}^{0}\rangle ,$
leading to that the solution state $|S\rangle $ could be found and the
unstructured search problem could be solved efficiently. However, as stated
below, there is a constraint on the power of an exponential
quantum-state-difference amplification. In fact, a unitary dynamical
state-locking process can be realized in polynomial time only for
polynomially many quantum states of the Hilbert space. It is not able to
transform simultaneously all the quantum states of the exponentially large
Hilbert space into their corresponding non-orthogonal states whose
differences can be made exponentially small in polynomial time. This
directly leads to that an exponential quantum-state-difference amplification
is not capable of distinguishing all the quantum states from one another in
the exponentially large Hilbert space. This constraint on the power of an
exponential quantum-state-difference amplification results in that one is
not able to use an exponential quantum-state-difference amplification to
distinguish unambiguously the unknown state $|\Psi _{k}^{S}\rangle $ from
the known state $|\Psi _{k}^{0}\rangle ,$ because the unknown state $|\Psi
_{k}^{S}\rangle $ is of the exponentially large search space. This leads to
that the unstructured search problem can not be solved efficiently, and this
result does not destroy the square speedup limitation. Whether or not the
exponentially large search space can be reduced to a polynomially small
state subspace (or subset) in polynomial time therefore is essential for a
polynomial-time unstructured quantum search process, because only when this
efficient search-space reduction can be realized via the quantum circuit
(2), can an exponential quantum-state-difference amplification be possibly
available for exponentially speeding up the quantum search process.

The unitary quantum dynamics has the unique ability to build the effective
interaction between the symmetric structure and property of the
tensor-product Hilbert space of a quantum system (or the quantum mechanical
principles) and the mathematical symmetric structure of a problem to be
solved on the quantum system (or the mathematical logic principles). It is
this effective interaction that can be used to realize efficiently the
information transfer from one form, i.e, the information of the solution
state as a whole to another, i.e., the information of the component states
of the solution state, and from the exponentially large search space of the
unstructured search problem to a polynomially small subspace (or subset) of
the tensor-product Hilbert space of the quantum system. Thus, the effective
interaction is essentially important to realize a polynomial-time quantum
search process. This fact reflects the inherent importance for the unitary
quantum dynamics to speed up a quantum computation. On the other hand, an
exponential quantum-state difference amplification, another part of the
polynomial-time quantum search process, is also constructed based on the
unitary quantum dynamics in time and space. Therefore, both the unitary
quantum dynamics and the symmetric structure and property of the
tensor-product Hilbert space of a quantum system are responsible for the
exponential speedup of a polynomial-time quantum search process. This
mechanism for exponential quantum-searching speedup also is very important
to understand the essence of exponential speedup of other efficient quantum
algorithms.\newline
\newline
{\Large 3. The unitary quantum dynamics and the universal quantum driving
force to speed up a quantum computation}

The unitary quantum simulation of nuclear spin systems or ensembles has been
studied extensively by the present author [36, 35, 37], which is related to
the author$^{\prime }$s research on the nuclear spin dynamics in the early
1990s. The unitary quantum dynamics has guided the author to construct the
quantum search processes in the past decade, although the constructed
quantum search process evolves from one form to another [21, 37, 29, 30]. It
also leads the author to discover the polynomial-time quantum search
processes [56]. As described above, it plays the central role in
constructing a polynomial-time quantum search process. Therefore, the
unitary quantum dynamics is closely related to a quantum-computing speedup.
However, the inherent importance for the unitary quantum dynamics to speed
up a quantum computation was not revealed until these extraordinarily fast
quantum algorithms [21, 32] are discovered by the present author in the
early 2000s (See also the comment [33]). Since then, the unitary quantum
dynamics including the time- and space-dependent one has been considered by
the author as the fundamental and universal principle to conduct research,
construction, and realization of a quantum computational process [21, 32,
40, 37, 29, 30, 41, 42, 43], although it has been suspected extensively that
the unitary quantum dynamics is a fundamental principle in macroscopic
world. The principle states explicitly that \textit{both a closed quantum
system and its quantum ensemble obey the same unitary quantum dynamics.}
Here the closed quantum system is referred to a pure-state system whose
density operator $\rho $ satisfies $Tr(\rho ^{2})=1,$ while its quantum
ensemble satisfies $Tr(\rho ^{2})<1.$ It should be pointed out that a
polynomial-time quantum search process [56] is constructed in a pure-state
quantum system and it has nothing to do with a quantum ensemble. But
according to this fundamental and universal principle the mechanism for
exponential quantum-searching speedup could be generally applicable not only
in a pure-state quantum system but also in a quantum ensemble. In turn, if
this exponential speedup mechanism is generally available [21, 32, 40, 33,
56] (See also the comment [48]), then this could indicate strongly that the
unitary quantum dynamics is the fundamental and universal principle not only
in a closed pure-state quantum system but also in a closed quantum ensemble
although it has been suspected extensively that the unitary quantum dynamics
is a fundamental principle in a macroscopic physical system [47]. This
problem will be further discussed later. Below it will be shown that the
unitary quantum dynamics in quantum mechanics is the universal quantum
driving force to speed up a quantum computation.

A universal quantum computational model [4, 5, 6, 7] obeys the three
fundamental principles (or properties): $(i)$ the unitary quantum dynamics, $%
(ii)$ the mathematical-logic principles, and $(iii)$ the quantum-state
description associated with a variety of fundamental quantum-state
properties. According to quantum mechanics [9] the first principle $(i)$ is
naturally compatible with the third one $(iii)$ in a quantum system to
execute a quantum computation. On the other hand, it may be considered that
the first principle $(i)$ is compatible with the second principle $(ii)$ in
the sense that every reversible mathematical-logic operation in a quantum
computation may be replaced with a discrete unitary operation. This
consideration is based on both the reversible classical computational models
[2] and the quantum Turing machine models [3]. However, there is a problem
that the second principle $(ii)$ may not be always compatible with the third
one $(iii)$ in a quantum computation in a universal quantum computational
model. The mathematical logic principles of a problem to be solved put a
constraint on the unitary quantum dynamics and also the possible application
of the quantum-state properties in a quantum computation. If in a quantum
computation one makes use of the quantum-state properties of a quantum
system which may be a pure-state system or a mixed-state ensemble, then
there could meet a conflict between the mathematical logic principles and
the quantum-state properties of the quantum system in the quantum
computation. In fact, such an inherent conflict is quite general in a
quantum computational process that needs to perform many different
reversible (mathematical-logic) functional operations in the Hilbert space
of a quantum system with a fixed number of qubits. For example, the unitary
manipulation on functional operations in a quantum computational process
needs to perform many reversible functional operations. Note that this
conflict is independent of any quantum measurement. Hence it is within the
universal quantum computational models under study at present. It is
different from the inherent conflicts early found [44, 45] due to the
quantum measurement in the halting operation in the universal quantum Turing
machine model [4]. However, all these conflicts inherent in the universal
quantum computational models can not lead to that one of the three
fundamental principles as stated above is more essential than the two other
principles in a quantum computation and especially in speeding up a quantum
computation. In particular, from these conflicts one is not able to deduce
that the unitary quantum dynamics is in the priority position in a universal
quantum computational model with respect to the mathematical-logic
principles and the fundamental quantum-state properties. \textit{Actually, a
universal quantum computational model itself is not able to decide which one
of these fundamental principles is more essential than the others in a
quantum computation. It is not yet able to resolve its inherent conflict
problems by itself.} Resolving these problems has to be based on the
quantum-computing speedup theory.

Here, using the quantum measurement to do a quantum computation is
considered to be beyond the universal quantum computational models under
study at present. It is not discussed in the paper. A universal quantum
computational model should be inherently consistent. One of the reasons why
here the quantum measurement is not included in these universal quantum
computational models is that too many conflicts will be generated among
these fundamental principles if the quantum measurement is added to these
universal quantum computational models. As an example, in the universal
quantum Turing machine model [4] both the unitary computational process and
the quantum parallel principle tend to conflict with the halting operation
[44, 30, 45] which is really involved in the quantum measurement. For
simplicity, the relativistic effect and the decoherence effect of a quantum
system are not yet considered in the paper.

A quantum-computing speedup theory studies mainly the mechanisms of
quantum-computing speedup and especially exponential quantum-computing
speedup. A quantum-computing speedup is used to measure the computational
power difference between quantum computation and classical computation when
both a quantum algorithm and the best classical algorithm solve a same
computational problem. A reasonable quantum-computing speedup theory should
satisfy some necessary conditions including: $(a)$ It is able to provide
reasonable mechanisms of exponential quantum-computing speedup for the
discovered polynomial-time quantum algorithms; $(b)$ It may lead to that a
universal quantum computational model is inherently consistent; $(c)$ It
should respect the fundamental and universal principle, i.e., the unitary
quantum dynamics; $(d)$ It is able to guide construction and realization of
a quantum algorithm and especially an efficient quantum algorithm. Here it
is required that the condition $(c)$ be satisfied mainly because the unitary
quantum dynamics plays the central role in constructing a polynomial-time
quantum search process. It also seems reasonable to require that the
quantum-computing speedup theory be consistent with the quantum parallel
principle in explaining the exponential speedup for the existing efficient
quantum algorithms. The conditions $(a)$ and $(c)$ are in the priority
position because a quantum-computing speedup theory studies mainly the
mechanisms of quantum-computing speedup. It seems reasonable and natural
that \textit{in the quantum-computing speedup theory the unitary quantum
dynamics is positioned at the center of a universal quantum computational
model}. However, it is impossible to show that the unitary quantum dynamics
is superior to the two other fundamental principles (or properties) within a
universal quantum computational model. The unitarity of the physical process
of a quantum computation is a necessary but not sufficient condition to set
up a universal quantum computational model, because unitarity of a pure
quantum physical process does not ensure that the mathematical-logic
principles are naturally satisfied in the process. The mathematical-logic
reversibility is also a necessary but not sufficient condition to set up a
universal quantum computational model, because it can only ensure the
reversibility of a computation in mathematical logic but not the unitarity
of the physical process of the computation. Therefore, a microscopic
physical process may obey the mathematical-logic principles or may not,
although it obeys the unitary quantum dynamics. From the point of view of
computational performance it seems that the quantum-state properties are
even more important for a quantum computation to outperform a classical
computation, and this has been supported by the investigation on the
computational performance of a universal quantum computational model in the
past two decades.

There are several important reasons why the unitary quantum dynamics should
be put at the heart of a universal quantum computational model in the
quantum-computing speedup theory. First of all, it is clear that such a
theory respects the fundamental and universal principle, i.e., the unitary
quantum dynamics and puts the principle in the priority position in a
universal quantum computational model. Next, if one puts the unitary quantum
dynamics at the heart of a universal quantum computational model, then the
conflict between the second $(ii)$ and third $(iii)$ principles as stated
above may be avoided. Obviously, there is not such a conflict for a pure
unitary quantum simulation. The conflict is met only when a quantum
computation is carried out. Now suppose that a computational problem is
solved by a quantum computational process on a quantum computer. Then the
mathematical logic principles that the problem obeys must be carried out in
a reversible or unitary form in the quantum computational process. Since the
unitary quantum dynamics is in the priority position, the quantum-state
properties that can be used in the computational process have to be suitably
constrained so that both the unitary quantum dynamics and the mathematical
logic principles are satisfied at the same time in the computational
process. This really means that there is not any conflict in the
computational process. Therefore, the quantum-computing speedup theory
immediately results in a fundamental rule to guide construction and
realization of a quantum algorithm in a universal quantum computational
model that a quantum computational process obeys the unitary quantum
dynamics and is compatible with the mathematical logic principles in a
reversible or unitary form. This fundamental rule has been used by the
present author to guide research, construction, and realization of a quantum
computational process in the past years [21a, 29]. It is also the basic
starting point of the present work. Another particularly important reason is
discussed below that is relevant to the realization of mathematical logic
principles in quantum computation.

A computation is to use a physical device (or computer) to treat a
computational task (or problem) according to a specific set of fundamental
mathematical-logic operation rules. In the computation the mathematical
logic principles that the problem obeys also are carried out by a specific
set of fundamental mathematical-logic operation rules. In the classical
Turing machine models [31] these fundamental mathematical-logic operation
rules are realized in irreversible form, while in the reversible classical
computational models [2] these fundamental operation rules are implemented
in reversible form. In these classical computational models these
fundamental operation rules have to be obeyed strictly during a computation
and hence these fundamental mathematical-logic operations are not allowed to
be manipulated. In the quantum Turing machine models [3] the fundamental
mathematical-logic operation rules are realized in unitary form. Just like
the reversible classical computational models, the quantum Turing machine
models do not allow these fundamental mathematical-logic operations to be
manipulated. Similarly, in the universal quantum computational models [4, 5,
6, 7] it is still required that these fundamental mathematical-logic
operation rules be strictly performed in reversible or unitary form. \textit{%
In fact, there is the most basic requirement for any computational model
that fundamental mathematical-logic operation rules be strictly and
correctly carried out in a computation}. This directly leads to that the
mathematical logic principles of a problem to be solved have to be satisfied
in a reversible or unitary form in a quantum computation in a universal
quantum computational model [21a, 29]. A universal quantum computational
model usually does not yet allow these fundamental mathematical-logic
operations to be manipulated. However, there is something more than the
fundamental mathematical-logic operation rules in a general quantum
computational process. There could exist a possibility to manipulate
unitarily the mathematical-logic operations of a problem to be solved in a
universal quantum computational model if purely quantum-mechanical
operations, excitations, and processes and so on are allowed to use in the
universal quantum computational model and the unitary manipulation does not
destroy the mathematical-logic principles that the problem obeys. Here these
purely quantum-mechanical operations, excitations, or processes need not
obey the fundamental mathematical-logic operation rules, but they obey
merely the quantum-mechanical principles. The quantum-computing speedup
theory also studies how these purely quantum-mechanical operations,
excitations, or processes manipulate the mathematical-logic operations to
speed up a quantum computation. It bridges the gap between quantum
computation and purely quantum-mechanical quantum simulation.

In general, mathematical-logic principles including mathematical symmetrical
structures of a problem to be solved do not affect essentially computational
performance, because they are independent on a detailed physical
computational model. A similar viewpoint also may be seen in Refs. [4, 5].
More intuitively, a computational problem is independent of a computer.
Thus, it is impossible that the mathematical-logic principles that the
problem obeys are able to affect essentially computational performance of
the computer. Though these mathematical logic principles can not affect
essentially the computational performance, the computational process to
solve the problem has to obey these mathematical logic principles [21a] no
matter whether the computer is classical or quantum. Thus, a
quantum-computing speedup to solve a problem makes sense only when the
mathematical-logic principles of the problem are first obeyed in a quantum
computational process. Of course, the mathematical-logic principles also are
obeyed in a classical computational process to solve the same problem.
However, here emphasizes the quantum computational process because a
quantum-computing speedup is determined alone by the quantum physical laws.
Why a quantum-computing speedup is determined only by the quantum physical
laws? Computational powers for the classical and quantum computers are
determined only by the classical and quantum physical laws, respectively.
Since the computational power for the best classical computational process
is unique, the quantum-computing speedup which is used to measure the
computational-power difference between quantum computation and classical
computation is only dependent on computational power of the quantum
computational process and hence it is determined alone by the quantum
physical laws. The precondition that a quantum-computing speedup makes sense
tends to be neglected in quantum computation. Researchers do not pay
attention to this precondition possibly because that fundamental
mathematical-logic operation rules are strictly and correctly carried out in
a computation is the most basic requirement for any computational model.
This precondition shows that whether or not a fundamental quantum physical
attribute (e.g., the unitary quantum dynamics, quantum superposition
principle, quantum coherence interference, and so on) is a universal quantum
driving force to speed up a quantum computation depends on whether or not
this fundamental attribute is able to perform independently and correctly
mathematical-logic operations in computation. Though the mathematical logic
principles do not affect essentially quantum computational performance, they
can determine which fundamental attribute in quantum mechanics may not be a
universal quantum driving force to speed up a quantum computation. On the
other hand, the polynomial-time quantum search processes [56] tell ones that
the effective interaction between the mathematical-logic principles and the
quantum physical laws plays a crucial role in achieving an essential speedup
for a quantum search process. This shows that only those fundamental
attributes in quantum mechanics that are able to build the effective
interaction may be considered as candidates of universal quantum driving
force to speed up a quantum computation. A fundamental quantum-physical
attribute may be a universal quantum driving force only if it can perform
independently and correctly mathematical logic operations in a quantum
computation and it is able to build the effective interaction between the
mathematical logic principles and the quantum physical laws. This is the
necessary condition that a fundamental attribute is able to become a
universal quantum driving force to speed up a quantum computation. It is
well known that these fundamental quantum-physical attributes including
quantum superposition principle, quantum coherence interference,
quantum-mechanical symmetry, the properties of quantum entanglement states,
the properties of multiple-quantum coherence mixed states, quantum
measurement, and so on, can not perform independently and correctly
mathematical logic operations in computation. Thus, all these fundamental
attributes are not a universal quantum driving force to speed up a quantum
computation. It is also impossible for the decoherence effect and nonlinear
effect (if it existed in a quantum system) to be a universal quantum driving
force to speed up a quantum computation. Note that the reversible classical
mechanics and equilibrium-state thermodynamics could perform independently
the mathematical logic operations. But because they can not build the
effective interaction between the mathematical logic principles of a problem
to be solved and the symmetric structure and property of the Hilbert space,
they are not a universal quantum driving force to speed up a quantum
computation. It is well known that the mathematical-logic principles can be
carried out independently and correctly by the unitary quantum dynamics. On
the other hand, the polynomial-time quantum search processes [56] tell ones
that the unitary quantum dynamics has the unique ability to build the
effective interaction between the mathematical logic principles and the
quantum physical laws. \textit{Therefore, the unitary quantum dynamics in
quantum mechanics is the universal quantum driving force to speed up a
quantum computation. }When people are talking about a quantum-computing
speedup in a quantum computation, they can not ignore the unitary quantum
dynamics that is restricted by the mathematical logic principles. Otherwise,
their obtained conclusion could make no sense! A typical example may be seen
in the comment [52]. These facts strongly support that the unitary quantum
dynamics has to be put at the center of a universal quantum computational
model.

It is very important that the unitary quantum dynamics in quantum mechanics
is the universal quantum driving force to speed up a quantum computation.
This result supports strongly that the mechanism for exponential speedup
deduced from the polynomial-time quantum search processes could be generally
available not only in a pure-state quantum system but also in a quantum
ensemble. In turn, this further implies that the unitary quantum dynamics is
the fundamental and universal principle not only in a pure-state quantum
system but also in a quantum ensemble, although it has been suspected
extensively that the unitary quantum dynamics is a fundamental principle in
a macroscopic physical system. Therefore, the quantum-computing speedup
theory supports strongly in theory that the unitary quantum dynamics is the
fundamental and universal principle in nature.

In the past two decades there are a large number of works to apply simply
the unitary time evolution processes (here relaxation or decoherence is
inevitable) of a physical system which may consist of spins, atoms,
molecules, or electrons, to implementing experimentally the quantum logic
gates, simple quantum algorithms, quantum simulations, and so on. These
works merely make use of the unitary quantum dynamics. They have little help
for studying and understanding the essence of the unitary quantum dynamics
to speed up a quantum computation. Understanding the unitary quantum
dynamics based on these works is not deeper and not yet more essential than
the one based on the conventional nuclear magnetic resonance (NMR)
experiments and electronic spin resonance (ESR) experiments (See also the
comment [33]). Therefore, these works have little help for understanding the
essential reason why the unitary quantum dynamics is considered as the
fundamental and universal principle in nature. For example, a
protein-folding process is a typical non-equilibrium process in nature. The
unitary quantum dynamics is destroyed strongly in the process. How can one
say the unitary quantum dynamics is a fundamental principle in such a
process? How can one say the unitary quantum dynamics governs such a
process? Only the mechanism for exponential quantum-searching speedup can
predict that \textit{such a process that does not obey the unitary quantum
dynamics is governed by the unitary quantum dynamics}.

As the central position of a universal quantum computational model the
unitary quantum dynamics could help to resolve the inherent conflicts
appearing in the universal quantum computational models. The inherent
conflicts [44, 30, 45] among the unitary quantum dynamics, the quantum
parallel principle, and the halting operation appear in the universal
quantum Turing machine model [4] due to the quantum measurement in the
halting operation. They include the conflict [44, 30] between the unitary
quantum dynamics and the quantum measurement and the conflict [44b, 30]
between the quantum parallel principle and the halting operation. There is
not the priority position for any one of the unitary quantum dynamics, the
quantum parallel principle, and the halting operation. In the paper [30] the
present author suggested that these conflict problems could be resolved on
the basis of the unitary quantum dynamics. This suggestion is based on the
fact that the unitary quantum dynamics is important in speeding up a quantum
computation. Now the unitary quantum dynamics is put at the center of the
universal quantum Turing machine model owing to its inherent importance to
speed up a quantum computation. Then the strategy in Ref. [30] to resolve
these inherent conflict problems becomes reasonable. A similar strategy also
may be adopted to resolve other conflict problems than those in Refs. [44,
30] in a universal quantum computational model, as described above.
Actually, a universal quantum computational model itself can not resolve its
inherent conflict problems. Resolving these inherent conflict problems must
be based on the quantum-computing speedup theory.

More importantly, the unitary quantum dynamics provides the fundamental
frame to realize the effective interaction between the mathematical-logic
principles of a problem to be solved and the quantum physical laws. A
universal quantum computational model also obeys the quantum-mechanical
symmetry. Quantum-mechanical symmetry of a quantum system that is different
from both the fundamental quantum-state properties and the unitary quantum
dynamics is a fundamental attribute of quantum world. The quantum-mechanical
symmetry under consideration here is mainly referred to the symmetric
structure and property of the Hilbert space of a composite quantum system.
It is based on the basic postulate in quantum mechanics that \textit{the
whole Hilbert space of a composite quantum system is a direct product (or
tension product) of the Hilbert spaces of component systems of the composite
system}. The quantum-mechanical symmetry includes the symmetric structure
and property of the multiple-quantum operator algebra spaces [35, 21a]. Of
course, there are also a number of the traditional quantum-mechanical
methods, i.e., the group-theory-based methods [9] (quantum-mechanical
symmetry is closely related to group theory) to characterize the symmetric
structure of the Hilbert space of a quantum system. But so far it has not
yet been shown that these traditional methods are useful to find an
efficient quantum algorithm in the frame of the quantum-computing speedup
theory. Here stress the multiple-quantum operator spaces because the
multiple-quantum transition processes are the basic quantum physical
processes in the quantum systems including multiple-spin systems. Global
symmetries of a closed quantum system such as the space translational
symmetry, rotational symmetry, and time displacement symmetry, which result
in the momentum, angular momentum, and energy conservation laws,
respectively, may not affect essentially quantum computational performance.
Of course, it does not rule out the possibility that the local versions of
these symmetries affect essentially quantum computational performance.
According to the quantum-computing speedup theory a quantum-mechanical
symmetry is not a universal quantum driving force to speed up a quantum
computation. However, in the frame of the unitary quantum dynamics the
symmetric structure and property of the Hilbert space may have an essential
effect on quantum computational performance. This is because it can affect
directly and most effectively the unitary quantum dynamics. The reason
behind it is that the symmetric structure and property of the Hilbert space
of a quantum system is closely related to the Hamiltonian of the quantum
system. The symmetric structure and property may make a direct impact on the
Hamiltonian of the quantum system, leading to that the effect of the
symmetric structure and property can be on the whole time evolution process
of the quantum system. On the other hand, though the mathematical logic
principles of a problem to be solved do not affect essentially quantum
computational performance, they can make a strong constraint on the unitary
quantum dynamics of the quantum system to solve the problem [21a]. This
leads to that the unitary quantum dynamics is able to bring together the
mathematical logic principles and the symmetric structure and property of
the quantum system and make them interacting with each other. As described
above, this interaction is of crucial importance to realize a
polynomial-time quantum search process.

The unitary quantum dynamics which acts as the fundamental frame also may
build the effective interaction between, on the one hand, the symmetric
structure and property of the Hilbert space and the time reversal symmetry
of a quantum system, and on the other hand, the mathematical logic
principles that a computational problem obeys. This leads to a new strategy
to construct an efficient quantum algorithm. A typical application of the
strategy may be seen in Refs. [21b, 40]. Here the strategy will not be
discussed in detail. It must be pointed out that all these methods or
strategies need to use the unitary manipulation on the functional operations
of the problem to be solved.\newline
\newline
{\Large 4. Quantum parallel principle and quantum-computing speedup}

The old quantum-computing speedup theory is based on the quantum parallel
principle [4]. The quantum parallel principle uses the quantum superposition
principle in quantum mechanics to speed up a quantum computation. It has an
intuitive feature similar to the classical parallel computation. The quantum
parallel principle has been considered to power essentially a quantum
computation in the past decades. It has been thought that this principle may
achieve an exponential quantum-computing speedup in some specific cases [11,
12, 13, 14, 15]. However, its generality is limited greatly. In the past
decades the quantum physical essence for how the quantum parallel principle
speeds up exponentially some specific quantum computations has been
investigated extensively. A vast number of works have been devoted to this
research subject. It has been thought extensively that the quantum
entanglement states could be responsible for the exponential
quantum-computing speedup [49]. However, a number of works [50] show that
there do not appear quantum entanglement states in the NMR experimental
tests of some quantum algorithms. A number of works also suspect or reject
that quantum entanglement states can help a quantum computation to achieve
an exponential quantum-computing speedup [51]. Some works show that an
exponential quantum-computing speedup could have nothing to do with quantum
entanglement states [40]. These works reflect mixed effect of quantum states
and their fundamental properties and especially quantum entanglement states,
multiple-quantum coherence mixed states, etc., and their fundamental
properties on a quantum computation.

There are some obstacles for quantum states of the Hilbert space of a
quantum system and their fundamental properties to become a quantum driving
force to speed up a quantum computation. According to the quantum-computing
speedup theory quantum states and their fundamental properties can not be a
universal quantum driving force to speed up a quantum computation, as
pointed out in the previous paragraph, because quantum states themselves and
their fundamental properties are not able to perform independently and
correctly mathematical logic operations in computation. Intuitively
speaking, a quantum computational process is a dynamical process, while
quantum states and their properties are static and passive and thus they can
not power independently a quantum computation. Moreover, quantum states
themselves are not able to distinguish a quantum computational process from
a purely quantum-mechanical dynamical process. Suppose that a black box
performs a unitary dynamical process. This process may be a purely
quantum-mechanical dynamical process or a quantum computational process that
is constrained by the mathematical-logic principles of a problem. Then input
and output quantum states of the black box are not able to determine whether
the unitary dynamical process is a purely quantum-mechanical process or it
is a quantum computational process (See the comment [52]). Note that the
intermediate quantum state in the process is completely determined by the
input and output quantum states for a given quantum algorithm. Therefore, it
is quite complex for quantum states and their fundamental properties to
affect a quantum-computing speedup. They could help the unitary quantum
dynamics to speed up a quantum computation. They could not have any
contribution to a quantum-computing speedup. Even they could have a negative
contribution to the quantum-computing speedup. \textit{If quantum states
themselves could speed up independently a quantum computation, then it would
be very possible that they sped up a non mathematical logic operation: }$%
1+2=5$\textit{. }

Though quantum states and their fundamental properties may not be a quantum
driving force to speed up a quantum computation, they could be able to help
the unitary quantum dynamics to speed up a quantum computation. Because
quantum states and their fundamental properties are independent of the
unitary quantum dynamics, their effect on mathematical-logic functional
operations usually is not effective. The effect of the quantum states tends
to be limited to the initial time of a functional operation. These are the
main reason why it tends to be ineffective for quantum states and their
fundamental properties to help the unitary quantum dynamics to achieve an
exponential quantum-computing speedup except for some special cases [11, 12,
13, 14, 15, 49c, 49d]. Among these fundamental quantum-state properties the
quantum superposition principle and quantum coherence interference could be
most useful to help the unitary quantum dynamics to speed up a quantum
computation.

In the past decades it has been considered extensively that the quantum
parallel principle leads to the exponential quantum-computing speedup for
the Abelian hidden-subgroup-problem (HSP) quantum algorithms [11, 12, 13,
14, 15] and some special non-Abelian HSP quantum algorithms [16]. It also
has been considered that quantum-computing speedup for the conventional
quantum search algorithms [23, 24, 25], quantum simulations [18], and other
quantum algorithms [49c, 49d] and so on originates from the quantum parallel
principle, simply since the quantum superposition states play the central
role or the quantum entanglement states appear in these quantum
computational (or simulation) processes [49b, 49d]. However, as shown above,
quantum states including quantum superposition states and quantum
entanglement states can not be a quantum driving force to speed up a quantum
computation. Here the quantum-computing speedup theory could provide
possible mechanisms for exponential speedup for these efficient HSP quantum
algorithms. More importantly, the theory could be helpful for improving
essentially these inefficient non-Abelian HSP quantum algorithms. As a
typical example, here gives a possible mechanism for exponential speedup of
an efficient HSP quantum algorithm on the basis of the unitary quantum
dynamics. An HSP quantum algorithm typically consists of the three parts:
the initial superposition state [4, 11, 7], a mathematical-logic functional
operation (See, for example, Refs. [13, 14, 15a, 53]), and the quantum
Fourier transform [14, 15c]. The basic structural characteristic feature for
an efficient HSP quantum algorithm is the mutual cooperation between the
last two parts in the frame of the unitary quantum dynamics and in the
suitable initial superposition state.

The mathematical symmetric structure of a hidden subgroup (HS) problem
constrains and modifies the unitary quantum dynamics of the quantum system
used to solve the problem. Then this constrained or modified unitary quantum
dynamics is applied to the suitable initial superposition state so that the
mathematical symmetric structure is mapped into the Hilbert space of the
quantum system. This is the first step of a usual HSP quantum algorithm.
This mapped symmetric structure in the Hilbert space may be called the
problem symmetric structure. It originates from the mathematical symmetric
structure of the HS problem alone, although it is in the Hilbert space. It
is different from the symmetrical structure of Hilbert space of the quantum
system itself. In Ref. [29] a cyclic group state space has a symmetric
structure similar to the current one, but that symmetric structure is
referred to that one of the Hilbert space of the quantum system itself. In
that case the search space of the search problem coincides with the cyclic
group state space (i.e., the Hilbert space). Now the information of solution
of the HS problem is hidden in the problem symmetric structure in the
Hilbert space. The second step of the HSP quantum algorithm is to apply a
suitable quantum Fourier transform to the quantum system so that the
solution information hidden in the problem symmetric structure could be
extracted in polynomial time from the Hilbert space by the quantum
measurement. It has been considered extensively that the second step is the
key step to achieve an exponential speedup for the HSP quantum algorithm. It
also has been recognized extensively that the problem symmetric structure in
the Hilbert space is key important for an efficient HSP quantum algorithm to
achieve an exponential speedup. Because the problem symmetric structure is
generated by the modified unitary quantum dynamics, the quantum Fourier
transform has to precisely cooperate with the modified unitary quantum
dynamics in the frame of the unitary quantum dynamics so that an exponential
speedup could be achieved in solving the HS problem. This is just the basic
structural characteristic feature for a polynomial-time HSP quantum
algorithm. Though sometimes the unitary quantum dynamics of the HSP quantum
algorithm could be locally destroyed, for example, some unitary operations
in the quantum algorithm could be replaced locally with the quantum
measurement processes [15b], this basic structural characteristic feature is
not changed essentially and hence the exponential speedup for the quantum
algorithm is still retained basically.

Because the problem symmetric structure in the Hilbert space originates from
the HS problem alone, it has nothing to do with the initial superposition
state. This is the main reason why the initial superposition state does not
have an essential effect on the exponential quantum-computing speedup to
solve the HS problem. This means that the exponential speedup is essentially
related to only the modified unitary quantum dynamics and the quantum
Fourier transform, although the initial superposition state is a necessary
component in an efficient HSP quantum algorithm.

The effect of the initial superposition state, which is the basic character
of the quantum parallel principle, on solving efficiently the HS problem
lies in that the superposition state helps the modified unitary quantum
dynamics to generate the problem symmetric structure in the Hilbert space in
polynomial time. In the old quantum-computing speedup theory the
superposition state is considered as the key to generating the problem
symmetric structure in polynomial time. But the new quantum-computing
speedup theory considers that the superposition state does not have any
ability to perform in a parallel form any mathematical-logic functional
operation, and only the modified unitary quantum dynamics is able to perform
independently and correctly the parallel mathematical-logic functional
operation. Therefore, the modified unitary quantum dynamics is really the
key to generating the problem symmetric structure in polynomial time,
although the initial superposition state is a necessary component in the
parallel functional operation.

This kind of exponential speedup to solve the HS problem is not general in
quantum computation. It is quite special because only in some special cases
the precise mutual cooperation is easy to realize between the modified
unitary quantum dynamics and the quantum Fourier transform in the suitable
initial superposition state. It is efficient to realize this precise mutual
cooperation for an Abelian HS problem [15]. However, it is generally hard to
realize the precise mutual cooperation for a non-Abelian HS problem [16]
except for some special cases. Different HS problem leads to different
modified unitary quantum dynamics. This further leads to that the precise
mutual cooperation needs to be re-realized so that an exponential speedup
could be re-achieved. If this mutual cooperation could no longer exist due
to that the HS problem is changed, then this exponential speedup could
disappear.

How can the quantum-computing speedup theory help to solve efficiently a
non-Abelian HS problem? Obviously, the quantum algorithm to solve the HS
problem does not consider explicitly the symmetrical structure and property
of the Hilbert space of the quantum system. This results in that the latter
does not have any effective effect on the problem symmetric structure in the
Hilbert space. From this point one may say that a usual HSP quantum
algorithm is not full quantum. This could be the main reason why this type
of efficient quantum algorithms are quite special. Therefore, there could be
a possible strategy [40, 21b] to improve these inefficient non-Abelian HSP
quantum algorithms that these inefficient quantum algorithms could be
improved essentially by making use of the symmetrical structure and property
of the Hilbert space of the quantum system. This strategy could be realized
only with the help of the unitary manipulation on the mathematical-logic
functional operations.\newline
\newline
{\Large 5. Discussion and conclusion}

In the past decade the present author has investigated extensively how the
fundamental quantum-mechanical principles and properties affect the quantum
computational performance of a quantum computer. Particular importantly, it
is proven that there are the polynomial-time unstructured quantum search
processes in quantum computation [56]. These works together form the
research basis of this paper. This paper is mainly devoted to studying the
mechanisms of quantum-computing speedup and especially exponential
quantum-computing speedup of a quantum computational process (or a quantum
algorithm). A quantum-computing speedup is determined only by the
fundamental quantum-mechanical principles and properties of a quanutm
system. It is independent on purely mathematical logic principles of a
problem to be solved on the quantum system. However, any one of these
fundamental quantum-mechanical principles and properties that is responsible
for speeding up a quantum computation has to obey these mathematical logic
principles. This results in that the interaction is inevitable between the
mathematical logic principles and the fundamental quantum-mechanical
principles. Such an interaction could have an essential effect on the
quantum computational process to solve the problem. These facts reflect
importance of the interaction to affect quantum computational performance.
At this point quantum computation is essentially different from classical
computation.

One of the mostly important results in the paper is to show on the basis of
the polynomial-time unstructured quantum search processes that the unitary
quantum dynamics in quantum mechanics is the universal quantum driving force
to speed up a quantum computation. A new quantum-computing speedup theory
therefore is set up on the basis of the unitary quantum dynamics in a
universal quantum computational model. It should be pointed out that both
the reversible classical computational models and the quantum Turing machine
models together are not sufficient to set up the theory. Both the unitary
quantum dynamics and the symmetric structure and property of the Hilbert
space of a quantum system are mainly responsible for an exponential
quantum-computing speedup for an efficient quantum algorithm. The inherent
importance for the unitary quantum dynamics to speed up a quantum
computation lies in the unique ability of the unitary quantum dynamics to
build the effective interaction between the symmetric structure and property
of the Hilbert space of a quantum system (or more generally the fundamental
quantum-mechanical principles) and the mathematical-logic principles
including the mathematical symmetric structure of a problem to be solved on
the quantum system. This unique ability is not owned by the reversible
classical mechanics and the reversible equilibrium-state thermodynamics. It
may generate an essential difference of computational power between quantum
computation and classical computation with the aid of the symmetric
structure and property of the Hilbert space. This effective interaction is
of crucial importance for a general quantum algorithm to achieve an
exponential speedup. In theory and experiment this effective interaction may
be built up with the help of the unitary manipulation on the
mathematical-logic functional operations of the problem.

It is well known that the classical computation has the three computing
resources: $(a)$ time or computational step number, $(b)$ space or memory,
and $(c)$ energy or precision. Quantum computation also has these three
computing resources. However, beside these three computing resources quantum
computation also has the computing resource that is not owned by the
classical computation. \textit{This quantum-computing resource is the
symmetric structure of Hilbert space of a composite quantum system}.

The old quantum-computing speedup theory is based on the quantum parallel
principle, while the latter is based on the quantum superposition principle
in quantum mechanics. Its underlying quantum-mechanical basis has been
extensively considered as the quantum entanglement states which are closely
related to the so-called quantum nonlocal effect. According to the new
quantum-computing speedup theory the existing (efficient) quantum algorithms
that are constructed in the frame of the old quantum-computing speedup
theory are not full quantum. These quantum algorithms, which include the
Abelian HSP quantum algorithms, non-Abelian HSP quantum algorithms, and
conventional quantum search algorithms and so on, have the common character
that the symmetric structure and property of the Hilbert space of the
quantum system to perform these quantum algorithms does not have any
effective effect on these quantum algorithms. This could be the main reason
why these efficient quantum algorithms are quite special and considered to
be semiclassical.

Since the early 2000s the unitary quantum dynamics has been considered as
the fundamental and universal principle that is responsible for speeding up
a quantum computational process in a physical system which may be a closed
pure-state quantum system or a closed quantum ensemble. Now this hypothesis
receives a strong support from the quantum-computing speedup theory. It is
based on the assertion that only the unitary quantum dynamics in quantum
mechanics is the universal quantum driving force to speed up a quantum
computation. This assertion really supports strongly that the unitary
quantum dynamics is the fundamental and universal principle not only in a
closed pure-state quantum system but also in a closed quantum ensemble. On
the other hand, it has been suspected extensively that the unitary quantum
dynamics is a fundamental principle to describe a non-equilibrium physical
process in a closed macroscopic physical system. Therefore, there is an
apparent conflict between the unitary quantum dynamics and a non-equilibrium
physical process in the physical system. How to resolve this apparent
conflict is a great challenge in physics. In Ref. [37] the present author
attempted for the first time to apply the quantum-computing speedup theory
(or its related idea) to describing a non-equilibrium physical process of a
closed physical system in the frame of the unitary quantum dynamics. This
idea was suggested first in a quantum spin system [37] and then in an atomic
physical system that has both the center-of-mass motion and internal motion
[30, 41]. However, solving ultimately this great challenge problem has a
long way to go.

The fundamental and universal principle that \textit{both a closed quantum
system and its quantum ensemble obey the same unitary quantum dynamics}
could be very helpful for understanding essentially a large number of
biophysical and biochemical processes in nature. It is well known that a
protein folding process is an ultrafast physical process in nature. A
classical search process generally can not match up the natural
protein-folding process with such an ultra-high folding speed. In contrast,
the polynomial-time quantum search process shows that such an ultrafast
physical process is understandable according to the quantum-computing
speedup theory. Moreover, the mechanism for exponential quantum-searching
speedup predicts that \textit{such an ultrafast natural process that does
not obey the unitary quantum dynamics could be governed by the unitary
quantum dynamics}.\newline
\newline
{\Large References}\newline
1. R. Landauer, \textit{Irreversibility and heat generation in the computing
process}, IBM J. Res. Develop. 5, 183 (1961)\newline
2. (a)\ C. H. Bennett, \textit{Logical reversibility of computation}, IBM J.
Res. Develop. 17, 525 (1973); (b) Y. Lecerf, \textit{Machines de Turing r%
\'{e}versibles}. \textit{R\'{e}cursive insolubilit\'{e} en }$n\in N$\textit{%
\ de l'\'{e}quation }$u=\theta ^{n}u,$\textit{\ o\`{u} }$\theta $\textit{\
est un isomorphisme de codes}, C. R. Acad. Sci. 257, 2597 (1963); (c) E.
Fredkin and T. Toffoli, \textit{Conservative logic}, Internat. J. Theor.
Phys. 21, 219 (1982)\newline
3. P. Benioff, \textit{The computer as a physical system: A microscopic
quantum mechanical Hamiltonian model of computers as represented by Turing
machines}, J. Statist. Phys. 22, 563 (1980); \textit{Quantum mechanical
Hamiltonian models of discrete processes that erase their own histories:
application to Turing machines}, Internat. J. Theor. Phys. 21, 177 (1982) 
\newline
4. D. Deutsch, \textit{Quantum theory, the Church-Turing principle and the
universal quantum computer}, Proc. Roy. Soc. Lond. A 400, 96 (1985)\newline
5. D. Deutsch, \textit{Quantum computational networks}, Proc. Roy. Soc.
Lond. A 425, 73 (1989)\newline
6. A. C. C. Yao, \textit{Quantum circuit complexity}, Proc. 34th Annual
Symposium on Foundations of Computer Science, IEEE Computer Society Press,
Los Alamitos, CA, PP. 352 (1993)\newline
7. E. Bernstein and U. Vazirani, \textit{Quantum computation complexity},
Proc. 25th Annual ACM Symposium on Theory of Computing, New York, PP. 11
(1993); Also see: SIAM J. Comput. 26, 1411 (1997)\newline
8. R. P. Feynman, \textit{Simulating physics with computers}, Internat. J.
Theor. Phys. 21, 467 (1982); \textit{Quantum mechanical computers}, Found.
Phys. 16, 507 (1986) \newline
9. L. I. Schiff, \textit{Quantum mechanics}, 3rd, McGraw-Hill book company,
New York, 1968\newline
10. (a) C. H. Bennett, \textit{Time / space trade-offs for reversible
computation}, SIAM J. Comput. 18, 766 (1989); (b) Y. Levine and A. T.
Sherman, \textit{A note on Bennett}$^{\prime }$\textit{s time-space tradeoff
for reversible computation}, SIAM J. Comput. 19, 673 (1990)\newline
11. D. Deutsch and R. Jozsa, \textit{Rapid solution of problems by quantum
computation}, Proc. Roy. Soc. London A 439, 553 (1992)\newline
12. A. Berthiaume and G. Brassard, \textit{The quantum challenge to
structural complexity theory}, Proc. 7th IEEE conference on structure in
complexity theory, Boston, PP. 132 (1992); Also see:\ \textit{Oracle quantum
computing}, J. Mod. Opt. 41, 2521 (1994)\newline
13. D. R. Simon, \textit{On the power of quantum computation}, Proc. 35th
IEEE Symposium on Foundations of Computer Science, Santa Fe, PP. 116 (1994);
Also see: SIAM J. Comput. 26, 1474 (1997)\newline
14. P. W. Shor, \textit{Polynomial-time algorithms for prime factorization
and discrete logarithms on a quantum computer}, SIAM J. Comput. 26, 1484
(1997); Also see: Proc. 35th Annual Symposium on Foundations of Computer
Science, IEEE Computer Society, Los Alamitos, CA, PP. 124 (1994)\newline
15. (a)\ A. Kitaev, \textit{Quantum measurements and the Abelian stabiliser
problem}, http://arxiv.org/abs/quant-ph/9511026 (1995); (b) R. Jozsa, 
\textit{Quantum algorithms and the Fourier transform}, Proc. R. Soc. Lond. A
454, 323 (1998); (c) R. Cleve, A. Ekert, C. Macchiavello, and M. Mosca, 
\textit{Quantum algorithms revisited}, Proc. R. Soc. Lond. A 454, 339
(1998); (d) M. Mosca, \textit{Quantum algorithms},
http://arxiv.org/abs/quant-ph/0808.0369 (2008); (e) A. M. Childs and W. van
Dam, \textit{Quantum algorithms for algebraic problems}, Rev. Mod. Phys. 82,
1 (2010)\newline
16. (a) M. Roetteler and T. Beth, \textit{Polynomial-time solution to the
hidden subgroup problem for a class of non-abelian groups},
http://arxiv.org/abs/quant-ph/ 9812070 (1998); (b) M. Ettinger, P. H$\phi $%
yer, and E. Knill, \textit{Hidden subgroup states are almost orthogonal},
http://arxiv.org/abs/ quant-ph/9901034 (1999); (c) S. Hallgren, A. Russell,
and A. Ta-Shma, \textit{Normal subgroup reconstruction and quantum
computation using group representations}, Proceedings of the 32nd Annual ACM
Symposium on Theory of Computing, PP. 627, Portland, Oregon, USA, 2000; (d)
R. Jozsa, \textit{Quantum factorization, discrete logarithms, and the hidden
subgroup problem}, http://arxiv.org/abs/quant-ph/0012084 (2000); (e) G.
Ivanyos, F. Magniez, and M. Santha, \textit{Efficient quantum algorithms for
some instances of the non-abelian hidden subgroup problem}, Internat. J.
Found. Comp. Sci. 14, 723 (2003); (f) M. Grigni, L. Schulman, M. Vazirani,
and U. Vazirani, \textit{Quantum mechanical algorithms for the nonabelian
hidden subgroup problem}, Proc. 33rd Annual ACM Symposium on Theory of
Computing, PP. 68, Crete, Greece, 2001; (g) C. Moore, D. Rockmore, A.
Russell, and L. J. Schulman, \textit{The hidden subgroup problem in affine
groups: basis selection in Fourier sampling},
http://arxiv.org/abs/quant-ph/0211124 (2002) \newline
17. (a) S. A. Cook, \textit{The complexity of theorem-proving procedures},
Proc. 3rd Annual ACM Symposium on Theory of Computing, PP. 151, New York,
USA, 1971; (b) R. M. Karp, \textit{reducibility among combinational problems}%
, In \textit{Complexity of Computer Computations}, PP. 85, Plenum Press, New
York, 1972; (c) L. A. Levin, \textit{Universal sequential search problems},
Problems of Information Transmission USSR 9, 265 (1973); (d) C. M.
Papadimitriou, \textit{Computational Complexity}, Addison-Wesley, Reading,
Massachusetts, 1994\newline
18. (a) S. Lloyd, \textit{Universal quantum simulators}, Science 273, 1073
(1996); (b) S. Wiesner, \textit{Simulations of many-body quantum systems by
a quantum computer}, http://arxiv.org/abs/quant-ph/9603028 (1996); (c) C.
Zalka, \textit{Simulating quantum system on a quantum computer}, Proc. R.
Soc. Lond. A 454, 313 (1998); (d) D. S. Abrams and S. Lloyd, \textit{A
quantum algorithm providing exponential speed increase for finding
eigenvalues and eigenvectors}, Phys. Rev. Lett. 83, 5162 (1999); (e) S.
Lloyd and S. L. Braunstein, \textit{Quantum computation over continuous
variables}, Phys. Rev. Lett. 82, 1784 (1999); (f) B. M. Terhal and D. P.
DiVincenzo, \textit{The problem of equilibration and the computation of
correlation functions on a quantum computer}, Phys. Rev. A 61, 22301 (2000)%
\newline
19. H. F. Trotter, \textit{On the product of semigroups of operators}, Proc.
Am. Math. Soc. 10, 545 (1959)\newline
20. M. Suzuki, \textit{Fractal decomposition of exponential operators with
applications to many-body theories and Monte Carlo simulations}, Phys. Lett.
A 146, 319 (1990); \textit{General theory of higher-order decomposition of
exponential operators and symplectic integrators}, Phys. Lett. A 165, 387
(1992) \newline
21. (a)\ X. Miao, \textit{Universal construction for the unsorted quantum
search algorithms}, http://arxiv.org/abs/quant-ph/0101126 (2001); (b) X.
Miao, \textit{Solving the quantum search problem in polynomial time on an
NMR quantum computer}, http://arxiv. org/abs/quant-ph/0206102 (2002) \newline
22. E. Knill, \textit{Approximation by quantum circuits},
http://arxiv.org/abs/quant-ph /9508006 (1995) \newline
23. L. K. Grover, \textit{Quantum mechanics helps in searching for a needle
in a haystack}, Phys. Rev. Lett. 79, 325 (1997)\newline
24. G. Brassard, P. H$\phi $yer, M. Mosca, and A. Tapp, \textit{Quantum
amplitude amplification and estimation},
http://arxiv.org/abs/quant-ph/0005055 (2000) \newline
25. E. Farhi, J. Goldstone, S. Gutmann, and M. Sipser, \textit{Quantum
computation by adiabatic evolution}, http://arxiv.org/abs/quant-ph/0001106
(2000) \newline
26. (a) C. H. Bennett, E. Bernstein, G. Brassard, and U. Vazirani, \textit{%
Strengths \ and \ weaknesses \ of \ quantum\ \ computing}, \
http://arxiv.org/abs/quant-ph/ 9701001 (1997); (b) M. Boyer, G. Brassard, P.
H$\phi $yer, and A. Tapp, \textit{Tight bounds on quantum searching},
Fortschr. Phys. 46, 493 (1998); (c) C. Zalka, \textit{Grover}$^{\prime }s$%
\textit{\ quantum searching algorithm is optimal}, Phys. Rev. A 60, 2746
(1999); (d) L. K. Grover, \textit{How fast can a quantum computer search?}
http://arxiv.org/abs/quant-ph/9809029 (1998)\newline
27. (a) W. van Dam, M. Mosca, and U. Vazirani, \textit{How powerful is
adiabatic quantum computation?}, http://arxiv.org/abs/quant-ph/0206003
(2002); (b) E. Farhi, J. Goldstone, S. Gutmann, and D. Nagaj, \textit{How to
make the quantum adiabatic algorithm fail}, http://arxiv.org/abs/0512159
(2005)\newline
28. (a) R. Beals, H. Buhrman, R. Cleve, M. Mosca, and R. De Wolf, \textit{%
Quantum lower bounds by polynomials}, Proceedings of 39th Annual Symposium
on Foundations of Computer Science, pp. 352 (1998); also see:
http://arxiv.org/ abs/quant-ph/9802049 (1998) and references therein; (b) E.
Farhi, J. Goldstone, S. Gutmann, and M. Sipser, \textit{Limit on the speed
of quantum computation in determining parity}, Phys. Rev. Lett. 81, 5442
(1998)\newline
29. X. Miao, \textit{Quantum search processes in the cyclic group state
spaces}, http:// arxiv.org/abs/quant-ph/0507236 (2005)\newline
30. X. Miao, \textit{The basic principles to construct a generalized
state-locking pulse field and simulate efficiently the reversible and
unitary halting protocol of a universal quantum computer},
http://arxiv.org/abs/quant-ph/0607144 (2006) \newline
31. A. M. Turing, \textit{On computable numbers, with an application to the
Entscheidungsproblem}, Proc. Lond. Math. Soc. S2-42, 230 (1937) \newline
32. X. Miao, \textit{A polynomial-time solution to the parity problem on an
NMR quantum computer}, http://arxiv.org/abs/quant-ph/0108116 (2001)\newline
33. The extraordinarily fast quantum-computing speed for these quantum
algorithms [21, 32] based on the unitary quantum dynamics also could be
achieved on the basis of the quantum parallel principle in nuclear spin
ensembles (See: R. Br\"{u}schweiler, Phys. Rev. Lett. 85, 4815 (2000)). It
could not show explicitly that the unitary quantum dynamics is more
essential than the quantum parallel principle in speeding up a quantum
computation, but it does show that the unitary quantum dynamics is at least
as powerful as the quantum parallel principle in solving extraordinarily
fast the same hard problems [21, 32] in nuclear spin ensembles. Therefore,
it reveals clearly that the unitary quantum dynamics is important to speed
up a quantum computation. Prior to these works [21, 32] the unitary quantum
dynamics had not been explicitly considered as a quantum driving force to
speed up a quantum computation. However, the inherent importance for the
unitary quantum dynamics to speed up a quantum computation has nothing to do
with whether the quantum system running a quantum algorithm is a pure-state
system or an ensemble. It lies in the unique ability of the unitary quantum
dynamics to build the interaction between the symmetric structure and
property of the Hilbert space of a quantum system and the mathematical
symmetric structure of a problem to be solved on the quantum system. That
the nuclear spin ensembles are able to reveal the importance is just due to
that a nuclear spin ensemble obeys the same unitary quantum dynamics as its
pure-state counterpart does. It is well known that a nuclear spin ensemble
with negligible spin relaxation effect obeys the unitary quantum dynamics
(See, for example, the NMR book [46a]) and so does an electronic spin
ensemble (See, for example, the ESR book [46b]).\newline
34. From the point of view of the unitary manipulation a space-independent
quantum system is usually a discrete quantum system that has only one
manipulating freedom degree of the discrete internal motion of the quantum
system. A spin system is a typical space-independent quantum system. On the
other hand, a single atom motioning in time and space may have two
independent manipulating freedom degrees of the discrete atomic internal
motion and the continuous or discrete atomic center-of-mass motion.
Therefore, a motional atom may not be a space-independent quantum system. 
\newline
35. X. Miao, \textit{Multiple-quantum operator algebra spaces and
description for the unitary time evolution of multilevel spin systems},
Molec. Phys. 98, 625 (2000)\newline
36. X. Miao, \textit{Universal \ construction \ of \ unitary \
transformation \ of \ quantum \ computation \ with \ one- and two-body \
interactions}, http://arxiv.gov/abs/ quant-ph/0003068 (2000)\newline
37. X. Miao, \textit{Efficient multiple-quantum transition processes in an }$%
n-$\textit{qubit spin system}, http://arxiv.org/abs/quant-ph/0411046 (2004)%
\newline
38. N. J. Cerf, L. K. Grover, and C. P. Williams, \textit{Nested quantum
search and NP-complete problem}, Phys. Rev. A 61, 032303 (2000)\newline
39. (a) L. K. Grover, \textit{Quantum computers can search rapidly by using
almost any transformation}, Phys. Rev. Lett. 80, 4329 (1998); (b) N.
Bhattacharya, H. B. van Linden van den Heuvell, and R. J. C. Spreeuw, 
\textit{Implementation of quantum search algorithm using classical Fourier
optics}, Phys. Rev. Lett. 88, 137901 (2002)\newline
40. X. Miao, \textit{A prime factorization based on quantum dynamics on a
spin ensemble (I)}, http://arxiv.org/abs/quant-ph/0302153 (2003)\newline
41. X. Miao, \textit{The STIRAP-based \ unitary \ decelerating \ and \
accelerating \ processes \ of a single free atom},
http://arxiv.org/abs/quant-ph/0707.0063 (2007)\newline
42. X. Miao, \textit{Unitarily manipulating in time and space a Gaussian
wave-packet motional state of a single atom in a quadratic potential field},
http: //arxiv.org/ abs/quant-ph/0708.2129 (2007) \newline
43. X. Miao, \textit{Unitary manipulation of a single atom in time and space
--- Spatially-selective and internal-state-selective triggering pulses},
unpublished\newline
44. (a) J. M. Myers, \textit{Can a universal quantum computer be fully
quantum?}, Phys. Rev. Lett. 78, 1823 (1997); (b) N. Linden and S. Popescu, 
\textit{The halting problem for quantum computers},
http://arxiv.org/abs/quant-ph/9806054 (1998); (c)\ M. Ozawa, \textit{Quantum
Turing machines: local transition, preparation, measurement, and halting},
http://arxiv.org/abs/quant-ph/9809038 (1998); (d) Y. Shi, \textit{Remarks on
universal quantum computer}, http://arxiv.org/abs/quant-ph /9908074 (1999);
Phys. Lett. A 293, 277 (2002) \newline
45. These conflicts do not exist in the universal quantum circuit model [5,
44]. Strictly speaking, they may be considered to be outside the universal
quantum computational models under study at present in the sense that they
really involve in the quantum measurement. In quantum measurement a quantum
system should be considered as an open system and hence its time evolution
process generally does not obey the unitary quantum dynamics. This is not
only because the quantum measurement results in collapse of the wave
function of the quantum system but also because there is a loss of
information of the quantum system to outside the system in the quantum
measurement.\newline
46. (a) R. R. Ernst, G. Bodenhausen, and A. Wokaun, \textit{Principles of
Nuclear Magnetic Resonance in One and Two Dimensions}, Oxford University
Press, Oxford, 1987; (b) A. Schweiger and G. Jeschke, \textit{Principles of
Pulse Electron Paramagnetic Resonance}, Oxford University Press, 2001\newline
47. L. D. Landau, E. M. Lifshitz, and L. P. Pitaevskii, \textit{Statistical
Physics (Part 1)}, 3rd edn, translated by J. B. Sykes and M. J. Kearsley,
Pergamon Press, New York, 1980\newline
48. That the mechanism for exponential quantum-searching speedup is
generally available does not mean that an exponential speedup always can be
achieved in a physical or computational process. However, if a physical or
computational process that is as hard as an NP-complete problem takes place
in polynomial time, then it is certain that the mechanism exists in the
process.\newline
49. (a) A. Ekert and R. Jozsa, \textit{Quantum algorithms:
entanglement-enhanced information processing}, Phil. Trans. Roy. Soc. Lond.
A 356, 1769 (1998); (b) R. Jozsa and N. Linden, \textit{On the role of
entanglement in quantum computational speed-up},
http://arxiv.org/abs/quant-ph/0201143 (2002); (c) E. Knill and R. Laflamme, 
\textit{On the power of one bit of quantum information}, Phys. Rev. Lett.
81, 5672 (1998); (d)\ S. Parker and M. B. Plenio, \textit{Efficient
factorization with a single pure qubit and }$\log _{2}N$\textit{\ mixed
qubits}, Phys. Rev. Lett. 85, 3049 (2000); Also see:
http://arxiv.org/abs/quant-ph/0102136 (2001) \newline
50. (a) S. L. Braunstein, C. M. Caves, R. Jozsa, N. Linden, S. Popescu, and
R. Schack, \textit{Separability of very noisy mixed states and implications
for NMR quantum computation}, Phys. Rev. Lett. 83, 1054 (1999); (b) T. M.
Yu, K. R. Brown, and I. L. Chuang, \textit{Bounds on the entanglability of
thermal states in liquid-state nuclear magnetic resonance}, Phys. Rev. A 71,
032341 (2005)\newline
51. (a) M. J. Bremner, C. Mora, and A. Winter, \textit{Are random pure
states useful for quantum computation?},
http://arxiv.org/abs/quant-ph/0812.3001 (2008); (b) D. Gross, S. T. Flammia,
and J. Eisert, \textit{Most quantum states are too entangled to be useful as
computational resources}, http://arxiv.org/abs/quant-ph/ 0810.4331 (2008);
(c)\ M. Van den Nest, \textit{Classical simulation of quantum computation,
the Gottesman-knill theorem, and slightly beyond}, http://arxiv.org
/abs/quant-ph/0811.0898 (2008)\newline
52. Here gives a testing program for the quantum-computing speedup of a
quantum algorithm. For a given efficient quantum algorithm according to the
qubit number, initial state, output state, and running time one may
construct a purely quantum-mechanical (QM) unitary sequence which is
equivalent to the quantum algorithm. Different qubit number may lead to
different initial state, output state, running time, and QM unitary
sequence. The testing program is carried out by two experimenters, in which
the quantum algorithm is executed in a black box. Experimenter One is
responsible for preparing the initial state and measuring the output state,
while Experimenter Two is responsible for putting the quantum algorithm or
its equivalent QM unitary sequence into the black box and setting the
running time. Experimenter Two need not tell Experimenter One which one of
the quantum algorithm and its equivalent QM unitary sequence is put into the
black box. For a given qubit number Experimenter One prepares the initial
state and then inputs it into the black box. Experimenter Two then may put
the quantum algorithm or its equivalent QM unitary sequence into the black
box. After the setting running time Experimenter One measures the output
state. If she or he finds that the output state is the desired state, then
she or he believes that the quantum algorithm indeed achieves the desired
quantum-computing speedup. This process may be repeated many times with
different qubit numbers, initial states, and running times till the final
result is obtained. Since sometimes Experimenter Two may put a purely QM
unitary sequence into the black box, this leads to that a wrong conclusion
that the purely QM unitary sequence also has an exponential speedup to solve
the same problem is obtained by Experimenter One from the measured result.
This is no sense! What is the reason? Because Experimenter One believes that
only the quantum states themselves are responsible for a quantum-computing
speedup, she or he need not care about which one of the quantum algorithm
and its equivalent QM unitary sequence is put into the black box, and what
she or he needs to care about is the input states and output states. This
special testing program above may be extended to a general case. As long as
Experimenter One does not consider explicitly every mathematical logic
operation of the efficient quantum algorithm, she or he could possibly
obtain a similar wrong conclusion. \newline
53. (a) D. Beckman, A. N. Chari, S. Devabhaktuni, and J. Preskill, \textit{%
Efficient networks for quantum factoring}, Phys. Rev. A 54, 1034 (1996); (b)
V. Vedral, A. Barenco, and A. Ekert, \textit{Quantum networks for elementary
arithmetic operation}, Phys. Rev. A 54, 147 (1996)\newline
54. An exponentially large space means that dimensional size of the space
increases exponentially with the qubit number of quantum system (or problem
size). A polynomially small space means that dimensional size of the space
increases polynomially with the qubit number.\newline
55. E. Farhi and S. Gutmann, \textit{An analog analogue of a digital quantum
computation}, Phys. Rev. A 57, 2403 (1998) or
http://arxiv.org/abs/quant-ph/9612026 (1996)\newline
56. X. Miao, \textit{a polynomial-time unstructured quantum search process
based on the unitary quantum dynamics and Hilbert-space symmetric structure}%
, unpublished\newline

\end{document}